\documentclass[aps,prc,floatfix,linenumbers]{revtex4}
\usepackage{amsmath,amssymb,amsfonts}
\usepackage{graphicx}% Include figure files
\usepackage{hhline}
\usepackage{bm}
\usepackage{amsmath}
\usepackage{epsfig}
\usepackage{graphicx}
\usepackage{multirow}

\usepackage{tabularx}
\usepackage{lscape}
\usepackage{rotating}
\usepackage{pstricks}

\usepackage{footnote}
\usepackage{hyperref}

\usepackage{pdfpages}
\usepackage{color} 

\usepackage{enumitem}

\newcommand{\bfvec}[1]{\hbox{\boldmath$#1$\unboldmath}}

\DeclareMathOperator{\erf}{erf}
\DeclareMathOperator{\erfc}{erfc}

\begin{document}
\title{Mechanical properties of the nucleon in the chiral confining model\\
II - in-medium evolution of the nucleon properties }

\author{Guy Chanfray$^{1}$, Hubert Hansen$^{1}$,  Bikram Keshari Pradhan $^{1}$}

\affiliation{$^{1}$ University Claude Bernard Lyon 1, CNRS/IN2P3, IP2I Lyon, UMR 5822, 69622  Villeurbanne, France \\
}

\begin{abstract}
This article is devoted to the study of the evolution of the properties of nucleons bound in nuclear matter within the framework of the chiral confining model. The in-medium nucleon trial states (either localized factorized wave functions or momentum-projected states) are determined by imposing the von Laue stability condition, according to the formal results established in a preliminary companion  paper (labeled as I) \cite{Nucleon-stability-I}.

The main results concern the response of the composite nucleon to the scalar field, as well as the respective roles of confinement and chiral symmetry breaking in the evolution of the in-medium nucleon mass. This evolution governs the repulsive three-body forces required for the nuclear saturation mechanism. We also analyze the modification of the energy density distribution and the pressure distribution inside the in-medium nucleon. We also draw some perspectives concerning the mapping between bound nucleon properties and the equation of state of dense matter as realized in the deep interior of neutron stars. 
\end{abstract}

\maketitle

%%%%%%%%%%%%%%%%%%%%%%%%%%%%%%%%%%%%%%%%%%%%%%%%%%%%%%%%%%%%%%%%%
%%%%%%%%%%%%%%%%%%%%%%%%%%%%%%%%%%%%%%%%%%%%%%%%%%%%%%%%%%%%%%%%%%%%%
\section{Introduction: Interplay between the nuclear many-body problem and nucleon structure: the chiral confining model}\label{Intro}
This article is the second part of a study that aims to investigate the properties of the nucleon, whether free or bound within the nuclear medium, using an approach that prioritizes a nucleon stability criterion, as detailed in the companion paper (labeled as I) \cite{Nucleon-stability-I}, where all formal developments are presented. The present approach aims to generalize these results to the case of a bound nucleon within a model in which massive constituent quarks are subject to a confining potential and are coupled to a surrounding pion cloud enveloping the quark core. A major objective is thus to calculate the response of the composite nucleon to the nuclear scalar field generated in a self-consistent manner, and thus to clarify the microscopic foundations of an approach aimed at developing a framework in which the properties of nuclear matter are linked, as much as possible, to the properties of the nucleon, and basic properties of low energy QCD, namely chiral symmetry breaking and color confinement. To properly assess the scope of this work, it is important to recall the origins of this theoretical framework.
\\

In order to bridge relativistic Walecka-type theories of nuclear matter \cite{SerotWalecka1986,Walecka1997} with chiral approaches, it was proposed in \cite{Chanfray2001} (see also \cite{Martini2006}) that the Walecka scalar meson $\sigma_W$, responsible for nuclear binding, can be identified with the chiral invariant scalar field $s=S-F_\pi$. The scalar field `s' is associated with the radial fluctuation of the chiral condensate $S$ around the ``chiral radius'' $F_\pi$, identified with the pion decay constant. 

\begin{figure}
\includegraphics[width=0.7\textwidth,angle=0]{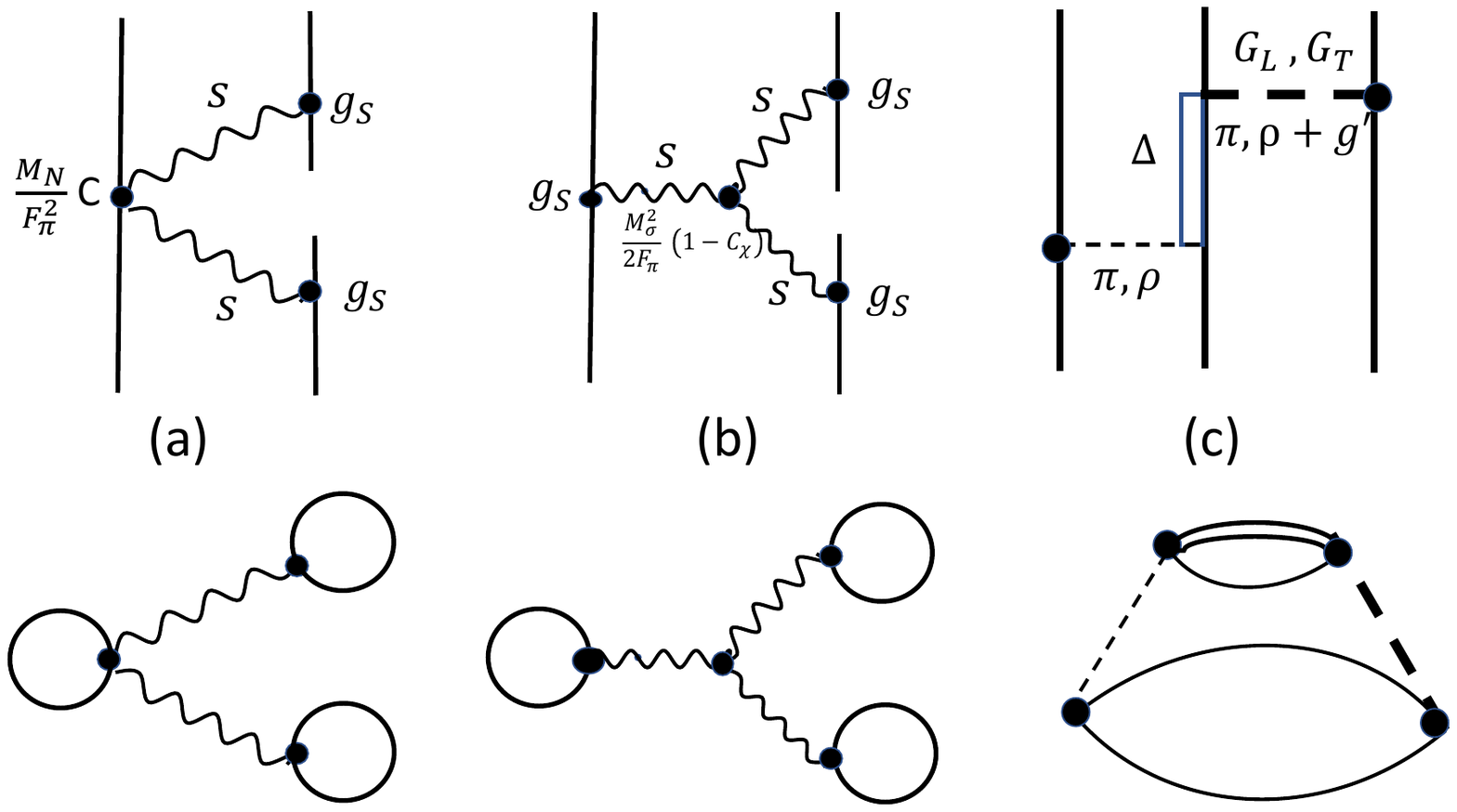}
\caption{Upper panel:  Nucleon three-body forces (3-hole-line) associated with the exchange of the scalar-isoscalar $s$ field in nuclear matter. (a) Response of the nucleon to the scalar field $s$ generating a repulsive three-body force; (b) tadpole diagram from the $s^3$ term in the chiral effective potential generating an attractive three-body force but corrected by the $C_\chi$ parameter; (c) another contribution to the three-body force presented in Ref.~\cite{Chanfray2024} coming from a Pauli-blocking effect in the $1-\Delta$ box (i.e., a medium modification of the $\sigma'$ exchange) corresponding to the earlier celebrated forces proposed in Refs. \cite{TB1,TB2,TB3,TB4,TB44} and also contained at NNLO in chiral EFT~\cite{CEFT}. Lower panel: the equivalent many-body diagrams. }
\label{THREEBODY}
\end{figure}

There is, however, a well-identified problem with nuclear saturation in the usual chiral effective theories \cite{Boguta83,KM74,BT01,C03}. Independent of the particular chiral model, in the nuclear medium, one can move away from the minimum of the vacuum effective potential (Mexican hat potential), i.e., into a region of smaller curvature. This single effect is equivalent to lowering the sigma mass and destroys stability, thereby undermining the applicability of such effective theories in the nuclear context. The effect can be associated with an $s^3$ tadpole diagram (see Fig.~\ref{THREEBODY}b), generating attractive three-body forces and destroying saturation, even if the repulsive three-body force from the Walecka mechanism is present.

These considerations led the development of a phenomenological model in the 2000s \cite{Chanfray2005,Chanfray2007,Ericson2007,Massot2008,Massot2009}, that we now call the chiral confining model (CCM), where it was proposed a  way to solve this problem through the  introduction of the nucleonic response to the scalar field, $\kappa_\mathrm{NS}$. This is also  the central ingredient of the quark--meson coupling model (QMC), introduced in the original pioneering work of P. Guichon \cite{Guichon1988} and successfully applied to finite nuclei with an explicit connection to the Skyrme force \cite{Guichon96,Guichon2004,Guichon2006,Stone} (see Ref.~\cite{QMCreview} for a recent review of this model). In that respect, the CCM model can be viewed as a specific chiral version of the QMC model. This effect of the reaction of the nucleon,   associated with the polarization of the quark substructure in the presence of the nuclear scalar field, will unavoidably generate three-body forces (see Fig.~\ref{THREEBODY}a), which may provide the desired repulsion to compensate the attractive tadpole. In practice, this response or, more precisely, the nucleon scalar  susceptibility $\kappa_\mathrm{NS}$ generates a non-linear coupling of the scalar field to the nucleon or, equivalently, a decrease in the scalar coupling constant, $g^*_S (s)=\partial M^*_N(s)/\partial s$, with increasing density.

The two specific main ingredients of the model are described below. 
\begin{itemize}
    \item 
 An in-medium modified nucleon mass is introduced, which is supposed to embed its quark substructure. This effective Dirac mass deviates from the bare nucleon mass in presence of the nuclear scalar field 
 $s$:
\begin{eqnarray}
		M^*_N(s)&=& M_N + g_S\, s +   \frac{1}{2}\kappa_{NS} \,s^2 + \mathcal{O}(s^3). 
  \end{eqnarray}
  The scalar nucleon coupling constant $g_S$ actually corresponds to the first order response of the nucleon to an external scalar field. The nucleon scalar susceptibility $\kappa_\mathrm{NS}$ is another response parameter which reflects the polarization of the nucleon, i.e., the self-consistent readjustment of the quark wave function in the presence of the scalar field.    Very generally the scalar coupling constant, $g_S$, and the nucleon response parameter,  $\kappa_{NS}$, depend on the subquark structure and the confinement mechanism as well as the effect of spontaneous chiral symmetry breaking. The estimation of these two parameters in a microscopic QCD-inspired approach, is the main objective of the present paper. In our previous works \cite{Chanfray2005,Chanfray2007,Massot2008,Massot2009,Massot2012,Rahul,Cham1,Chanfray2023,Universe,Chanfray2024,Cham2} we introduced a dimensionless parameter, 
\begin{equation}
 C\equiv \frac{\kappa_\mathrm{NS}\,F_\pi^2}{2 M_N},   
\end{equation}
which is expected to be of the order $C\sim 0.5$ as in the MIT bag  used in the QMC framework. 
\item
In the majority of our previous works \cite{Chanfray2005,Martini2006,Chanfray2007,Massot2008,Massot2009,Massot2012,Rahul,Cham1}, the chiral effective potential governing the dynamics of the nuclear scalar field $s$ had the simplest linear sigma model (L$\sigma$M) form. However for reasons given in Ref. \cite{Chanfray2023,Universe,Chanfray2024,Cham2}, we recently used an enriched chiral effective potential from  the Nambu-Jona-Lasinio (NJL) model able to give a correct description of the low-energy realization of chiral symmetry in the hadronic world. It is defined by the Lagrangian given in Eq.~(9) of Ref.~\cite{Chanfray2023} or Eq.~(59) of Ref.~\cite{Universe}. As it was established in Ref.~\cite{Chanfray2023}, the net effect of the use of the NJL model is to replace the L$\sigma$M chiral potential by its NJL  counterpart which is very well approximated by its expansion to third order in $s$:
\begin{equation}
V_{\chi,\mathrm{NJL}}(s)= \frac{1}{2}\,M^2_\sigma\, {s}^2\, +\,\frac{1}{2}\,\frac{M^2_\sigma -M^2_\pi}{ F_\pi}\, {s}^3\,\big(1\,-\,C_{\chi,\mathrm{NJL}}\big) +...\, .\label{vchiNJL} 
\end{equation}
For a given pion mass, sigma mass and pion decay constant parameters, the main difference with the original phenomenological version using the L$\sigma$M  lies in the presence of the $C_{\chi,\mathrm{NJL}}$ parameter whose expression in terms of a NJL loop integral is given in Ref.~\cite{Chanfray2023}. Considering typical values of the NJL parameters, we obtain values is in the range $C_{\chi,\mathrm{NJL}}: 0.4-0.5$.  The effect of the NJL model through the specific parameter $C_{\chi,\mathrm{NJL}}$ is thus to reduce the attractive tadpole diagram in the chiral potential, which makes it more repulsive.
\end{itemize}
Combining the effects of the tadpole diagram  (i.e, associated with the $s^3$ in the chiral potential, (see Fig.~\ref{THREEBODY}~b) and of the response parameters (see Fig.~\ref{THREEBODY}~a), it is possible to show that the scalar sector generates a three-nucleon contribution to the energy per nucleon, 
\begin{equation}
E^{(3b-s)}
=\frac{g^3_S}{2\,M^4_\sigma\,F_\pi}\,\left(2\, \frac{M_N}{g_S\,F_\pi}\,C -\,\left[1\,-\,C_\chi\right]\right)\,\rho^2_s\equiv\frac{g^3_S}{2\,M^4_\sigma\,F_\pi}\,\left(2\, \left[\frac{M_N}{g_S\,F_\pi}\,C +\,\frac{1}{2}\,C_\chi\right]\,-\,1\right)\,\rho^2_s,
\label{THREEBODN}
\end{equation}
where $\rho_s$ is the nucleonic scalar density, see Refs.~\cite{Ericson2007,Rahul,Chanfray2023,Chanfray2024,Universe,Cham2} for more details. One very important point is that  these scalar response parameters can be related to some chiral properties of the nucleon. According to the lattice data  analysis of the Adelaide group, \cite{LTY03,LTY04,TGLY04,AALTY10}, the nucleon mass can be expanded in terms of the square of the pion mass, $M^2_{\pi}$, as $M_N(M^2_{\pi}) = 
a_{0}\,+\,a_{2}\,M^2_{\pi}\, +\,a_{4}\,M^4_{\pi}\,+ ...+\,\Sigma_{\pi}(M^2_{\pi},\, \Lambda)$ 
 where the pionic self-energy, $\Sigma_{\pi}(M^2_{\pi}, \Lambda)$, contains the non analytical contribution and is explicitly separated out. The explicit connection between the lattice QCD parameters $a_2$ and $a_4$, and the response parameter $g_S$ and $C$ in the L$\sigma$M case, has been first given in Refs.~\cite{Chanfray2007,Ericson2007,Massot2008}, and in its more complete form in our recent papers  \cite{Rahul,Cham1,Cham2,Universe,Chanfray2023,Chanfray2024}: 
\begin{equation}
a_2= \frac{F_\pi\, g_{S}}{M^2_{\sigma}}, 
\quad
a_4 = \frac{F_\pi\,g_{S}}{2 M^4_{\sigma }}\,\left(2\,\frac{M_N}{g_S\,F_\pi}\,C \,-\,3 \left[1\,-\,C_\chi\right]\right)\equiv\frac{F_\pi\,g_{S}}{2 M^4_{\sigma }}\,\left(2\,\left[\frac{M_N}{g_S\,F_\pi}\,C +\,\frac{3}{2}\,C_\chi\right]\,-\,3 \right).\label{LATTIX}
\end{equation}  
 Depending of the details of the chiral extrapolation, the extracted values of the parameters are within the range in between  $a_2^\simeq 1.5$~GeV$^{-1}$, $a_4\simeq -0.5$~GeV$^{-3}$~\cite{LTY04} and $a_2\simeq 1.0$~GeV$^{-1}$, $a_4\simeq -0.25$~GeV$^{-3}$~\cite{LTY03,AALTY10}. The quantity $a_2 M^2_\pi \sim 20-30$~MeV represents the non pionic piece of the sigma commutator directly associated with the scalar field $s$ (see the detailed discussion of this quantity in Ref.~\cite{Chanfray2007}). One very robust conclusion is that the lattice result for $a_4$ is much smaller than the one obtained in  the simplest linear sigma model ignoring the nucleonic response ($C=0$) and the enriched form of the chiral potential ($C_\chi=0$), for which $a_4\simeq -3.5$~GeV$^{-3}$. Hence  agreement with lattice data requires a strong compensation from effects governing the three-body repulsive force needed for the saturation mechanism: compare Eqs.~(\ref{LATTIX}) and (\ref{THREEBODN}). In addition the compatibility of lattice data, nucleon modeling and saturation properties absolutely requires a non vanishing value of this $C_\chi$ parameter~\cite{Chanfray2023,Universe,Chanfray2024,Cham2}.
 
 This model has been  applied in the past to the equation of state of nuclear matter~\cite{Chanfray2005,Chanfray2007,Massot2008,Massot2009,Rahul,Chanfray2024,Universe,Cham1,Cham2} and neutron stars~\cite{Massot2008,Massot2012,Cham1,Cham2} as well as to the study of chiral properties of nuclear matter~\cite{Chanfray2005,Martini2006,Ericson2007,Chanfray2007} at different levels of approximation in the treatment of the many-body problem (RMF, Relativistic Hartree-Fock or RHF, pion loop correlation energy). Among other results, one important lesson is the importance of a coherent treatment of the Fock term including the rearrangement terms \cite{Massot2008,Cham1}  which has to be consistently incorporated at the level of the self-energies and total energy. Although this problem is of minor importance for what concerns the binding energy around nuclear matter density, the use of a Hartree-Fock basis in place of the Hartree basis for the nucleon Dirac wave function, may play a very important role at high density in limiting the maximum mass of hyperonic neutron star as pointed out in Ref. \cite{Massot2012}. A short review is given in a recent paper~\cite{Chanfray2024} where in addition an improved calculation of the correlation energy is presented together with the idea of the ``two sigma meson" picture ($s$ and $\sigma'$) each of them being  associated with  three-body forces; this later point is summarized in Fig.~\ref{THREEBODY} \\
\\
 In parallel with these phenomenological developments, works aimed at providing a more microscopic foundation for the CCM model has been carried out. First and foremost, we can cite Ref.~\cite{Chanfray2011} where an explicit construction of the background scalar field $\mathcal{S}$, associated with the radial fluctuation mode of the chiral condensate, was performed in the NJL model using a bosonization technique based on an improved derivative expansion valid at low (space-like) momenta~\cite{Chan}. The vacuum expectation value of the chiral invariant scalar field $\mathcal{S}$ coincides with the constituent quark mass $M_0\sim 350\,\mathrm{MeV}$ and the  ``nuclear physics" scalar field $S$  is recovered by rescaling the chiral invariant scalar field $\mathcal{S}$, according to :
\begin{equation}
 \mathcal{S}=\frac{M_0}{F_\pi}\,S=\frac{M_0}{F_\pi}\left(s+F_\pi\right).   
\end{equation}
 The important point is that its fluctuating piece, i.e., the  $s$ field, has to be identified with the usual ``nuclear physics sigma meson" of relativistic Walecka theories, $\sigma_W$. 
 Various confining interactions have been incorporated (quark--diquark string interaction, linear and quadratic confining interaction) on top of the NJL model to generate not only the enriched NJL potential but also values for the response parameters $g_S$ and $\kappa_{NS}$. In two recent papers \cite{Universe,Universe2025} we have proposed an original method whose aim is to provide a theoretical framework to pave the way between QCD phenomenology and nuclear physics phenomenology.   It is based on the field correlator method (FCM) developped by Y. Simonov and collaborators \cite{Simonov1997,Simonov1998,Tjon2000,Simonov2002,Simonov2002a,Simonov-light,light1,light2,Digiacomo,Simonov2019}. This idea has been originally implemented in details in  \cite{Universe}.
By performing a gluon averaging in the euclidean QCD partition function, on can generate an effective non perturbative interaction between quarks mediated by a gluon correlator parametrized in a convenient gaussian form:
\cite{Tjon2000,Simonov2002}: 
\begin{equation}
D(x)=\frac{\sigma}{2\pi T^2_g} \, e^{-x^2/4 T^2_g}\,\,\,\,\hbox{with}\,\,\,\,\,T_g=\sqrt{\frac{9\sigma}{\pi^3 \mathcal{G}_2}}.
\end{equation}
It depends on two  QCD quantities measured in lattice QCD \cite{Digiacomo}, namely the string tension, $\sigma=0.18$~GeV$^2$, and the gluon correlation length, $T_g=0.25 - 0.3\, fm$, itself related to the gluon condensate $\mathcal{G}_2$.  As explained in detail 
in section 3.3 of Ref. \cite{Universe}, modulo some ansatz prescription, this approach allows  to generate simultaneously, at a semi-quantitative level,  a confining interaction built from the gluon correlator with long distance ($R\gg T_g$)  behaviour $V_C(r)\sim\sigma \, r$,    together with an equivalent NJL model  with scalar interaction strength $G_1=120\pi\sigma T^4_g/(4N_cN_f)\sim 10$~GeV$^{-2}$ and cutoff $\Lambda\sim 1/T_g\sim 600$~MeV. In the most recent paper \cite{Universe2025}, the entire theoretical framework linking the phenomenology of QCD and the phenomenology of nuclear matter was discussed. This paper ends with a chapter where the properties of the nucleon and its evolution in nuclear matter are calculated, along with an estimation of the nuclear response parameters, $g_S$ and $\kappa_{NS}$ (or $C$). It included not only the effect of the string-like confining potential, but also the effect of the pion–quark coupling (with the pionic sector Lagrangian derived from the bosonization of the underlying NJL model). However, the very delicate issue of the mechanical stability of the nucleon in the presence of both a confining potential and a surrounding pionic cloud was not addressed in a fully satisfactory manner. This point is, in fact, the main motivation of the present article.
\\

The rest of the paper is organized as follows. In the  section~\ref{sec:Nucleon_in_CCM}, we present how trial states for in-medium nucleons can be constructed. Starting from the effective action derived within the FCM framework, which combines the coupling of quarks to the pion field with a string-like confining interaction, we present the effective Lagrangian governing the structure of the (in-medium)  nucleons, whose parameters depend on the nuclear scalar field. In  section~\ref{sec:EMT_CCM_PaperI}, by generalizing the results of the companion paper I \cite{Nucleon-stability-I} to the case of the in-medium nucleon, we present how an effective Hamiltonian and an energy–momentum tensor (EMT) can be constructed, including a nonlocal coupling of the pionic field to the constituent quarks. We also recall how the von Laue condition for mechanical stability is formulated, which in practice allows us to fix the size parameter $b$ governing the nucleonic trial Fock-space amplitude. Based on paper I,  section~\ref{sec:STATIC_EMT_CCM} briefly recalls how the static EMT tensor can be derived and how internal spatial distributions of energy and pressure can be defined for an in-medium nucleon. Section ~\ref{sec:STATIC_LOCALIZED_N} presents the results obtained for a localized nucleon state, where the product of three quark orbitals gives the nucleon wave function (or the Fock-space amplitude), and where the center of the confining force coincides with the center-of-mass (CM) position, identified with the three-quark string junction position. In section~\ref{sec:MOMENTUM_PROJECTED_N}, we present the results obtained for a nucleon state better suited to describe a physical nucleon, namely a state properly projected onto a well-defined three-momentum. From this, one can deduce the evolution of the mass and size (isoscalar rms radius), as well as other nuclear properties, and obtain a microscopic prediction for the scalar coupling constant $g_S$ and the dimensionless scalar susceptibility parameter $C$. In section ~\ref{sec:NUCLEON_NS}, we discuss perspectives concerning a possible mapping between nucleon structure properties (i.e., the relation between the internal mechanical pressure and energy density) and the equation of state of dense uniform matter, as realized in the deep interior of neutron stars. We conclude our findings in the section ~\ref{sec:CONCLUSION}.  In order to keep the paper as self-contained as possible, the main formal results derived in the companion paper I \cite{Nucleon-stability-I} are summarized in the appendix.
%%%%%%%%%%%%%%%%%%%%%%%%%%%%%%%%%%%%%%%%%%%%%%%%%%%%%%%%%%%%%%%%%%%%%%%%%%%%%%%
%%%%%%%%%%%%%%%%%%%%%%%%%%%%%%%%%%%%%%%%%%%%%%%%%%%%%%%%%%%%%%%%%%%%%%%%%%%%%%%%%%%%%
\section{The In-medium nucleon in the chiral confining model}\label{sec:Nucleon_in_CCM}
%%%%%%%%%%%%%%%%%%%%%%%%%%%%%%%%%%%%%%%%%%%%%%%%%%%%%%%%%%%%%%
%%%%%%%%%%%%%%%%%%%%%%%%%%%%%%%%%%%%%%%%%%%%%%%%%%%%%%%%%%%%%

Let us now start with the core of the work, namely applying the formal developments of the companion paper I \cite{Nucleon-stability-I}to the case of the in-medium nucleon.

\subsection{Construction of the (in-medium) nucleon states }
In the chiral confining model (CCM) the (in-medium) color singlet nucleon state (ignoring for simplicity its explicit spin-color-flavor structure) is generically represented as ($V$ is a normalization volume):
\begin{equation}
\left|N(\Phi)\right\rangle = V^{3/2}\,\int \frac{d^3 p_1}{(2\pi)^3}\frac{d^3 p_2}{(2\pi)^3}\frac{d^3 p_3}{(2\pi)^3}
\,\Phi\left({\bf p}_1, {\bf p}_2, {\bf p}_3\right)\,B^\dagger_{{\bf p}_1}(\mathcal{S}) B^\dagger_{{\bf p}_2}(\mathcal{S}) B^\dagger_{{\bf p}_3}(\mathcal{S}) \left|\varphi(\mathcal{S})\right\rangle, \label{NUCWF}
\end{equation}
\begin{equation}
\int \frac{d^3 p_1}{(2\pi)^3}\frac{d^3 p_2}{(2\pi)^3}\frac{d^3 p_3}{(2\pi)^3}\,|\left|\Phi\left({\bf p}_1, {\bf p}_2, {\bf p}_3\right)\right|^2 =1.\label{NORM1} 
\end{equation}
From the normalization condition on the Fock-space amplitude, $\Phi\left({\bf p}_1, {\bf p}_2, {\bf p}_3\right)$, this state is automatically  normalized to 1 ($\left\langle N\right| \left|N\right\rangle =1$) in a large box of volume $V$.
As explained in section 5 of our review paper \cite{Universe2025} (see also the other recent papers \cite{Universe,Chanfray2024}), $\left|\varphi(\mathcal{S})\right\rangle$ represents a BCS-like chirally broken vacuum constructed in the framework of an underlying NJL model. Hence $\left|\varphi(\mathcal{S})\right\rangle$ is the vacuum of the creation operators $B^\dagger_{{\bf p}_1}(\mathcal{S})$, (i.e., $B_{{\bf p}_1}(\mathcal{S})\left|\varphi(\mathcal{S})\right\rangle=0$) of constituent quarks of mass   $M_q=\mathcal{S}\equiv \mathcal{S}(\bfvec{R}_N)$, this mass being identified with the value of the nuclear  scalar field felt by the in-medium nucleon, whose center of mass position is $\bfvec{R}_N$. All  the quantities related to the nucleon properties  should be seen as a functional of the scalar field $\mathcal{S}$. In the spirit of the Born-Oppenheimer approximation this value of $\mathcal{S}$ coincides with the value of the scalar field $\mathcal{S}(\bfvec{R}_N)=(M_0/F_\pi)(F_\pi + s(\bfvec{R}_N))$ taken at the nucleon center of mass (CM) position, $\bfvec{R}_N$, hence assuming that the variation of the scalar field  is slow enough to neglect its variation in the nucleonic volume. This reasoning  is presented in full in various QMC papers \cite{Guichon96,QMCreview}. In our approach inspired from the field correlator method (FCM) \cite{Simonov1997,Simonov1998,Tjon2000,Simonov2002a,Simonov2002,Simonov-light,light1,light2,Digiacomo,Simonov2019},  this nucleonic volume is  $V_N\sim 1/\sigma^{3/2}$, with  $\sigma \simeq 0.18\,GeV^2$, being  the string tension.  The chiral phase associated with this BCS-like vacuum is such that:
\begin{equation}
\sin\varphi(p ;\mathcal{S})=\frac{{\mathcal{S}}}{E_p({\mathcal{S}})}\equiv\frac{{\mathcal{S}}}{\sqrt{p^2 +{\mathcal{S}}^2}}.
\end{equation}
The bare nucleon is recovered when the scalar field $\mathcal{S}$ coincides with the vacuum constituent quark mass, $M_0$, solution of the gap equation \cite{Universe2025,Universe}:
\begin{equation}
M_0 = m\,+\,4N_c N_f M_0\,G_1\,I_1(M_0)\quad\hbox{with}\quad  I_1(\mathcal{S})=\int_0^\Lambda \frac{d{\bf p}}{(2\pi)^3}\,\frac{1}{2\,E_p(\mathcal{S})}.  \label{GAP}
\end{equation}
The quark field operator  for a given flavor and  color can be expanded on the BCS basis, according to:
\begin{equation}
q({\bf{r}},t)=\frac{1}{\sqrt{V}}\,	\sum_k\,\eta_k\,C_k(t)\,e^{i\,\bf{k}\cdot\bf{r}}.\label{QEXP}
\end{equation}
The $C_k$'s are positive and negative energy state annihilation operators representing either a true destruction operator of a constituent quark or a true creation operator of an antiquark.
The states labeled by  $k=({\bf{k}}, s, \varepsilon)$  correspond to a plane wave basis: $\bf{k}$ is the  momentum, $s=\pm 1$ represents the polarization state and $\varepsilon$ allows to distinguish between positive and negative energy states:
\begin{eqnarray}
& &\varepsilon=+1,\quad C_{\bf{k}, s, +1}=B_{\bf{k}, s}\, ,\quad \eta_{\bf{k}, s, +1}=u(\bf{k},  s)\nonumber\\
& &\varepsilon=-1,\quad C_{\bf{k}, s, -1}=D^\dagger_{-\bf{k}, -s}\, ,\quad \eta_{\bf{k}, s, -1}=v(-\bf{k}, -s).
\end{eqnarray} 
The explicit forms of the the Dirac spinors in term of the  chiral angles ($s_k=\sin\varphi_k, c_k=\cos\varphi_k$) are: 
\begin{equation}
\eta_{\bf{k}, s, +1}=u({\bf k}, s)=\sqrt{\frac{1}{2}}\left(
\begin{array}{c}
                \sqrt{1+s_k}\,\,\chi_s \\
                \sqrt{1-s_k}\,\,\bfvec{\sigma}\cdot {\bf k}\,\,\chi_s
\end{array}
\right)
\end{equation}
\begin{equation}
\eta_{\bf{k}, s, -1}=v(-{\bf k}, -s)=\sqrt{\frac{1}{2}}\left(
\begin{array}{c}
                -\sqrt{1-s_k}\,\,\bfvec{\sigma}\cdot {\bf k}\,\,\chi_s \\
                \sqrt{1+s_k}\,\,\chi_s 
\end{array}
\right).
\end{equation}
%%%%%%%%%%%%%%%%%%%%%%%%%%%%%%%%%%%%%%%%%%%%%%%%%%%%%%%%%%%%%

\subsection{The nucleon trial states, localized CCM static bag, physical CCM nucleon, and model parameters}
The general picture underlying our approach can be summarized as follows: nuclear
matter is made of (possibly in-medium modified ) nucleons that look like Y-shaped strings, which are generated by a non perturbative
confining force, with constituent quarks (with mass $M_q=\mathcal{S}$)  at the ends moving in the (possibly modified) chirally broken vacuum.
In the following we  first consider the case of an object where the center of mass (CM) position $\bfvec{X}_N$, identified with the three-quark string junction position, is fixed. We name such an object,  ``localized CCM static bag" denoted by $|N(\bfvec{X}_N)\rangle$. In this case   the three-quark Fock-space amplitude  is just the product of the three single-quark Fock-space amplitudes properly projected onto the color singlet state with $I=J=1/2$, namely,
\begin{eqnarray}
\Phi_{\bfvec{X}_N}\left({\bf p}_1, {\bf p}_2, {\bf p}_3\right)&=&e^{-i\,(\bfvec{p}_1 +\bfvec{p}_2+\bfvec{p}_3)\cdot\bfvec{X}_N}\,\, \Phi\left({\bf p}_1\right)\,  
\Phi\left({\bf p}_2\right)\,  
\Phi\left({\bf p}_3\right)\\
\Phi\left({\bf p}_1\right)\,  
\Phi\left({\bf p}_2\right)\,  
\Phi\left({\bf p}_3\right)&=&R_0(p_1) R_0(p_2) R_0(p_3)\label{WFNOCM},
\end{eqnarray}
with the normalization condition: 
\begin{equation}
\int \frac{d^3 p}{(2\pi)^3}\,\left|\Phi\left({\bf p}\right)\right|^2 =1.\label{NORM2}   
\end{equation}
In practice we will use a  gaussian trial single-quark Fock-space amplitude:
\begin{equation}
\Phi(\bfvec{p})= R_0(p)= (2\pi)^{\frac{3}{2}}\,\left(\frac{b^2}{\pi}\right)^{\frac{3}{4}}\,e^{-b^2\,p^2/2} \label{TRIALWF0}.  
\end{equation}
A significant  source of improvement in the description of nucleon properties   is to replace the localized nucleon state, not possessing a well-defined total 3-momentum, by a momentum projected nucleon state with well defined momentum $\bfvec{P}$, denoted by $|N(\bfvec{P}_N)\rangle$, or sometimes for short  by $|\bfvec{P}_N\rangle$ in the formal developments. As in Ref. \cite{Gastao}, this amounts to enrich the factorized Fock-space amplitude by the appropriate projection according to: 
\begin{equation}
\Phi_{\bfvec{P}_N}\left({\bf p}_1, {\bf p}_2, {\bf p}_3\right)=\frac{1}{V^{1/2}} N_{\bf{P}_N} \Phi\left({\bf p}_1\right)\,
\Phi\left({\bf p}_2\right)\,  
\Phi\left({\bf p}_3\right)\,(2\pi)^3\,\delta \left(\bf{P}_N - \bf{p}_1 -\bf{p}_2 - \bf{p}_3 \right)\label{WFCM},
\end{equation}  
where $N_{\bf{P}_N}$ is a normalization factor, whose explicit expression is 
$$N_{\bf{P}_N}=\left(\frac{3}{4\pi b^2}\right)^{3/4}\,\exp\left(\frac{P^2_N\, b^2}{6}\right),$$ 
when using the gaussian trial ansatz of Eq.~\eqref{TRIALWF0}. In the notation $\bfvec{P}_N$, the subscript $N$ indicates that the state is normalized and carries momentum $\bfvec{P}$. Hence, it is implicitly understood that $\bfvec{P}_N\equiv\bfvec{P}$.\\
In the following we will use the following notations:
\begin{equation}
|N: \bfvec{0}_X\rangle\,\equiv\,|N(\bfvec{X}_N=0)\rangle,\qquad    |N: \bfvec{0}_P\rangle\,\equiv\,|N(\bfvec{P}_N=0)\rangle .\label{NOTATION}
\end{equation}
For the numerical calculations will  we use $\sigma=0.18\,GeV$, $T_g=0.283\,fm$ (or $\mathcal{G}_2=0.025\,GeV^4$), $M_0=356.7\,MeV$, $M_\pi\equiv M_\pi^{vac}=140\,MeV$, $F_\pi\equiv F_\pi^{vac}=91.7\,MeV$, issued from  the NJLset1  parameter set given in in our previous review paper \cite{Universe2025}. The parameter $c_\pi$, governing the pion size (see below, Eq.\eqref{PIONSIZE} is the only remaining free parameter. We will compare our results with the accepted values  of the nucleon axial coupling constant, $g_A=1. 26$, and of the mean square radius, $\sqrt{\langle r^2_B\rangle}$, of the quark core baryon number distribution, itself related to the the isoscalar radius of the nucleon, $\sqrt{\langle r^2_S\rangle}\simeq 0.77\, fm$ \cite{Xiong}, according to: 
\begin{equation}
  \sqrt{\langle r^2_B\rangle}=\sqrt{\langle r^2_S\rangle\,-\,\frac{6}{m^2_\omega}} =0.46\,fm.
\end{equation}
%%%%%%%%%%%%%%%%%%%%%%%%%%%%%%%%%%%%%%%%%%%%%%%%%%%%%%%%%%%%%%%%%
%%%%%%%%%%%%%%%%%%%%%%%%%%%%%%%%%%%%%%%%%%%%%%%%%%%%%%%%%%%%%
\subsection{The in-medium  effective action and the effective Lagrangian}
According to \cite{Universe2025} the appropriate effective action for the description of the (in-medium) nucleon containing both the effect of the confining potential and the leading order coupling of the pion field $\vec{\Phi}$ to the constituent quark reduces to:
\begin{eqnarray}
S_F &=& \int d^4x\,\bar{q}(x)\left[i\,\gamma^{\mu}\partial_\mu-{\cal S}\,+\,\frac{1}{2\,F_\pi({\cal S})}\gamma^j\gamma^5\partial_j\vec{\Phi}\cdot\vec{\tau}\right](x)q(x)\,-\,\frac{1}{2}M_\pi^2(\mathcal{S})\,\vec{\Phi}^2 \,+\, \frac{1}{2}  \,\partial^{\mu}\vec{\Phi}\cdot\partial_{\mu}\vec{\Phi}\nonumber\\
&&-\,
\int d^4x\, d^4y\,\bar{q}(x)M({\bf x}, {\bf y})\,\delta(x_0-y_0) q(y)\nonumber\\
&&\qquad\hbox{with}\qquad F^2_\pi(\mathcal{S})=\mathcal{S}^2\,\tilde{I} (\mathcal{S})=\mathcal{S}^2\,2 N_c N_f\,\tilde I_{2}({\cal S})=\mathcal{S}^2\,\frac{2 N_c N_f I_2({\cal S})}{1\,+\,4\,G_2\,{\cal S}^2\,2 N_c N_f\,I_2({\cal S})},\nonumber\\
&&\qquad M^2_\pi(\mathcal{S})=\frac{m \mathcal{S}}{G_1 F^2_\pi(\mathcal{S})}, \qquad I_2(\mathcal{S})=\int_0^\Lambda \frac{d{\bf p}}{(2\pi)^3}\,\frac{1}{4\,E^3_p(\mathcal{S})}.
\end{eqnarray}
The coupling constants $G_1$ (scalar) and $G_2$ (vector), the current quark mass $m$, and the cutoff  $\Lambda$ are the parameters entering the underlying NJL model described at length in \cite{Universe} and \cite{Chanfray2011}. This in-medium effective action differs from the vacuum one mainly by the replacement of the vacuum pion mass and the vacuum pion decay constant with the in-medium modified quantities $M_\pi(\mathcal{S})$ and $F_\pi(\mathcal{S})$ given above. The effect of the inclusion of the confining kernel induced by the presence of the string junction, is embedded in the  static mass operator, $M({\bf x}, {\bf y})$, whose explicit form is given in Eqs. (120,121) of Ref. \cite{Universe2025}: 
\begin{eqnarray}
M({\bf x}, {\bf y})&=&\int\frac{d^3 k}{(2\pi)^3} e^{-i {\bf k}\cdot ({\bf x} - {\bf y})}\, \gamma_0\,\frac{s_k- c_k\,\bfvec{\gamma}\cdot{\bf{k}}}{2}\,\gamma_0\,\left(W_C\left({\bf X}\right)+ W_C\left({\bf Y}\right)\right). 
\end{eqnarray}
The effective on-body  confining potential is such that,
\begin{eqnarray}
 W_C\left({\bf R}\right)&=&  \frac{\sigma}{2\,\sqrt{\pi}\,T_g}\,{\bf R}^2
\,\int_0^1 dv\,\int_0^1 dw\,e^{-\left(\left(v-w\right)^2\frac{{\bf R}^2}{4 T^2_g}\right)}\,-\,\frac{2\,\sigma\,T_g}{\sqrt{\pi}},\label{VCONF}\nonumber\\
&=&\sigma\left(R -\frac{2\,T_g}{\sqrt{\pi}} +\frac{2\,T_g}{\sqrt{\pi}}  e^{-\left(\frac{R^2}{2\,T^2_g}\right)}\,-\, R\,\erfc\left(\frac{R}{2 T_g}\right)\right)\,-\,\frac{2\,\sigma\,T_g}{\sqrt{\pi}}, \label{CONFPOT}
\end{eqnarray}
where $\sigma$ is the string tension and $T_g$ is the gluon correlation length, itself related to the gluon condensate $\mathcal{G}_2$ according to $T_g=\sqrt{9\sigma/\pi^3 \mathcal{G}_2}$. With the aim of addressing later the virial theorem and the von Laue condition for the mechanical stability of the (isolated or in-medium) nucleon, we introduce what could be called a “confining pressure potential” which turns out to have a very simple expression:
\begin{equation}
 \bfvec{r}\cdot \bfvec{\nabla}_{\bfvec{r}}W_C(\bfvec{r})=r\,W'_C(r)\,= \,\sigma\,r\,\erf\left(\frac{r}{2 T_g}\right). \label{PRESSPOT}
\end{equation}\\
If we project the quark field onto the positive energy sector (ignoring constituent $q\bar{q}$ in the Fock-state amplitude), i.e. if we limit its decomposition  on the positive energy BCS states and use the property $U_{\bf p}\eta_{\bf p}\equiv(\beta\, \sin\varphi_p\,+\,\,\bfvec{\alpha}\cdot\hat{\bf p}\,\cos\varphi_p)\,\eta_{\bf p}=\eta_{\bf p}$, one easily demonstrates that the confining piece of the action is equivalent to a local static Lagrangian according to:
\begin{equation}
  -\,\int d^4x d^4y\,\bar{q}(x)M({\bf x}, {\bf y})\,\delta(x_0-y_0) q(y) = -\,\int d^4x\, \bar{q}(x)\,\gamma^0\,
  W_C\left({\bf x}\right)q(x)\equiv \int d^4x\, \mathcal{L}_C(x).
\end{equation}
It follows that the dynamics of the in-medium bound quark, in the presence of both the string potential and the pion field, is described by the following local Lagrangian:
\begin{eqnarray}
 \mathcal{L}(\mathcal{S}) &=& \mathcal{L}_{\pi Q}(\mathcal{S})\,+\,\mathcal{L}_C\nonumber\\
 &=& \bar{q}(x)\left[i\,\gamma^{\mu}\partial_\mu-{\cal S}+\frac{1}{2 F_\pi(\mathcal {S})}\gamma^j\gamma^5\partial_j\vec{\Phi}\cdot\vec{\tau}\right](x)q(x)
 \,-\,\frac{1}{2}M_\pi^2(\mathcal{S})\,\vec{\Phi}^2 \,+\, \frac{1}{2}  \,\partial^{\mu}\vec{\Phi}\cdot\partial_{\mu}\vec{\Phi}\nonumber\\
 &&\,-\,q^\dagger(x)\,
  W_C\left({\bf x})\,\right)q(x).\label{LAGEFF}
\end{eqnarray}
Let us stress again that ${\cal S}$ represents the value of the nuclear scalar field taken  at the nucleon center of mass (CM) position, $\bfvec{R}_N$ and which is supposed to be uniform  in the nucleonic volume in the spirit of the Born-Oppenheimer approximation. The isolated nucleon case is simply recovered when $\mathcal{S} =M_0$.\\
We thus recover the generic Lagrangian given by Eq. (1) of Paper I \cite{Nucleon-stability-I}, but with the values of the generic parameters replaced by scalar-field–dependent parameters:
\begin{equation}
 M_q \rightarrow  \mathcal{S}, \qquad F_\pi\rightarrow F_\pi(\mathcal {S}),\qquad M_\pi\rightarrow M_\pi(\mathcal {S}).
\end{equation}
All formal results concerning the total energy (the nucleon mass), the mean pressure, the energy density, and the pressure distribution inside the nucleon, established in paper I, can be fully reproduced by making these replacements. For completeness, all results actually used for the numerical calculations are listed in the Appendix.
%%%%%%%%%%%%%%%%%%%%%%%%%%%%%%%%%%%%%%%%%%%%%%%%%%%%%%%%%%%%%%%%%%%%%%%%%%%%%%%%%%%%%%%%%%%%%%%%%%%%%%%%%%%%%%%%%%%%%%%%%%%%%%%%%%%%%%%%%
\section{Hamiltonian, Energy momentum tensor and von laue stability condition for the in-medium nucleon state}\label{sec:EMT_CCM_PaperI}
\subsection{The Hamiltonian: quark coupling to finite size pions}
A Hamiltonian can be immediately derived from the Lagrangian \eqref{LAGEFF}. However, as was discussed and justified in Paper I \cite{Nucleon-stability-I}, we were led to replace the local coupling of the pion field to the quark source  with a non-local coupling that takes into account the finite size of the pion, thereby avoiding a collapse of the quark core. This modified  Hamiltonian therefore  reads:
\begin{eqnarray}
H &=& H_Q\,+\,H_{\pi Q}\,+\,H_{\pi\pi} \label{HTOT}\\   
&\equiv&\int d^3 r\,\left(\mathcal{H}_Q\,+\,\mathcal{H}_{\pi Q}\,+\,\mathcal{H}_{\pi\pi} \right)(\bfvec{r})\\
&\equiv&\int d^3 r\,\left(\mathcal{H}_{QK}\,+\,\mathcal{H}_{QM}\,+\,\mathcal{H}_{C}\,+\,\mathcal{H}_{\pi Q}\,+\,\mathcal{H}_{\pi c} 
\,+\,\mathcal{H}_{\pi K}\,+\,\mathcal{H}_{\pi M}\right)(\bfvec{r})\label{HDENS}\\
H_Q &=& \int d^3 r\, q^\dagger(\bfvec{r})\left[-i\bfvec{\alpha}\cdot\bfvec{\nabla}_{\bfvec{r}}\,+\, \beta \mathcal{S} \, +  \,\,W_C(\bfvec{r})\right]\,q(\bfvec{r})\equiv H_{QK}\,+\, H_{QM}\,+H_C\label{HQ}\\
H_{\pi Q} &=&- \int d^3 r\,\frac{1}{2 F_\pi({\cal S})}\,\partial_j\vec{\Phi}(\bfvec{r})\cdot \bar{q}(\bfvec{r})\gamma^j\,\gamma_5\,\vec{\tau}q(\bfvec{r})\equiv 
-\int d^3 r\,\frac{1}{2 F_\pi({\cal S})}\,\bfvec{\nabla}\vec{\Phi}(\bfvec{r})\cdot \tilde{\bfvec{S}}(\bfvec{r})\label{HPIQ}\\
H_{\pi\pi}&=& \int d^3 r\,\frac{1}{2}\left(\vec{\Pi}^2\,+\,\partial_{j}\vec{\Phi}\cdot\partial_{j}\vec{\Phi}\,+\, M_\pi^2(\mathcal{S})\,\vec{\Phi}^2\right)(\bfvec{r})\equiv H_{\pi c}\,+\,H_{\pi K}\,+\,H_{\pi M}.\label{HPI}
\end{eqnarray}
In the expression of $H_{\pi Q}$, the operator $\tilde{\bfvec{S}}(\bfvec{r})$ represents  a delocalized quark source for the pion field,  
\begin{eqnarray}
  \tilde{\bfvec{S}}(\bfvec{r})= \int d^3x \,\tilde{P}(\bfvec{r}-\bfvec{x})\vec{\bfvec{S}}(\bfvec{x})=\int d^3x \,\int \frac{d^3t}{(2\pi)^3}\,e^{-i{\bf t}\cdot({\bf r}-{\bf x})}\,P({\bfvec{t}})\,\vec{\bfvec{S}}(\bfvec{x}),  
 \end{eqnarray}
 where  the original quark source, $\vec{\bfvec{S}}(\bfvec{r})$, is given below along with its longitudinal Fourier transform:
\begin{equation}
\vec{\bfvec{S}}(\bfvec{r})= \bar{q}(\bfvec{r})\bfvec{\gamma}\,\gamma_5\,\vec{\tau}q(\bfvec{r}),\quad  
 \vec{S}^\dagger_{\bf k} =\int d^3r \,e^{-i{\bf k}\cdot{\bf r}} \,\vec{\bfvec{S}}(\bfvec{r})\cdot{\hat{\bf k}},\quad  
 \vec{S}_{\bf k} =\int d^3r \,e^{i{\bf k}\cdot{\bf r}} \,\vec{\bfvec{S}}(\bfvec{r})\cdot{\hat{\bf k}}.
\end{equation}
Accordingly the longitudinal spin-isospin operator is modified according to:
\begin{eqnarray}
 \vec{S}_{\bf k} =\int d^3r \,e^{i{\bf k}\cdot{\bf r}} \,\vec{\bfvec{S}}(\bfvec{r})\cdot{\hat{\bf k}}&\to&
\tilde{S}_{\bf k}= \int d^3r \,e^{i{\bf k}\cdot{\bf r}} \,\tilde{\bfvec{S}}(\bfvec{r})\cdot{\hat{\bf k}}=P (\bfvec{k})\,\vec{S}_{\bf k}.
\end{eqnarray}
The spreading function $\tilde{P}(x)$ describes the coupling between quarks and a pion of finite spatial extent. As discussed in detail in our previous works, \cite{Universe2025} and  I \cite{Nucleon-stability-I}, the introduction of a finite pion size, of the order of the quark core size itself,  provides a resolution to a long-standing issue encountered in this class of chiral bag models \cite{Brown,Pirner}. Indeed, if one retains an elementary pion field,  the model, as it stands, suffers from an instability problem due to  a diverging attractive pion-quark interaction energy (i.e., the pionic self-energy) when the quark core size goes to zero. As the size of the quark core decreases, the negative pion pressure may become very large surpassing the Fermi pressure and potentially causing  collapse of the ``quark core bag".\\
We take for $\tilde{P}(\bfvec{\eta})$ a  gaussian form:
$$\tilde{P}(\bfvec{\eta})=\left(\frac{1}{\pi\,\rho^2_\pi(\mathcal{S})}\right)^{3/2}\,\exp\left(-\frac{\eta^2}{\rho^2_\pi(\mathcal{S})}\right),\qquad P({\bf{t}})=\exp\left(-\rho^2_\pi(\mathcal{S})\frac{t^2}{4}\right).$$
We expect the size parameter $\rho_\pi$ to be directly related to the electromagnetic size of the pion. Since this pion radius arises from a constituent quark loop \cite{KLE}, we also expect it to depend on the in-medium constituent quark mass, $M=\mathcal{S}$, and consequently on the in-medium pion decay constant parameter $F_\pi(\mathcal{S})$ \cite{Pirner}. For orientation we  take:
\begin{equation}
\rho^2_\pi(\mathcal{S})= \frac{c_\pi}{4\pi^2 F^2_\pi(\mathcal{S})}=\frac{F^2_\pi}{F^2_\pi(\mathcal{S})}\,c_\pi\,(0.34\, fm)^2\label{PIONSIZE}.
\end{equation}
%%%%%%%%%%%%%%%%%%%%%%%%%%%%%%%%%%%%%%%%%%%%%%%%%%%%%%%%%%%%%%%%%%%%%%%%%%%%%%%%%%
\subsection{Local energy-momentum tensor (EMT) in presence of a non local pion-quark coupling}
Although the finite-size effect of the pion–quark coupling makes the Lagrangian non-local, we showed in section III-A  of the companion  paper I \cite{Nucleon-stability-I} that it is still possible to define a local energy–momentum tensor, formally of the same form as the one derived from the local Lagrangian, but with $\vec{\bfvec{S}}(\bfvec{r})$ replaced by $\tilde{\bfvec{S}}(\bfvec{r})$.\\
The components of the EMT tensor, which are relevant for our study are :
\begin{eqnarray}
 \hat{T}^{00}(x)&=&\left(\mathcal{H}_{QK}\,+\,\mathcal{H}_{C}\,+\,\mathcal{H}_{\pi Q}\,+\,\mathcal{H}_{\pi c} 
\,+\,\mathcal{H}_{\pi K}\,+\,\mathcal{H}_{\pi M}\right)(x)\\
\hat{P}(x)=\frac{1}{3}\hat{T}^{ii}(x)&=&\left(\frac{1}{3}\mathcal{H}_{QK}\,+\,\frac{1}{3}\mathcal{H}_{\pi Q}\,+\,\mathcal{H}_{\pi c} 
\,-\,\frac{1}{3}\mathcal{H}_{\pi K}\,-\,\mathcal{H}_{\pi M}\right)(x).
\end{eqnarray}
$\hat{T}^{00}$  coincides exactly with the Hamiltonian density introduced before where the various partial Hamiltonian density operators are given by \eqref{HTOT},\eqref{HDENS},\eqref{HQ},\eqref{HPIQ},\eqref{HPI}. The operator $\hat{P}(x)$ can be referred as the local  pressure  operator, in the sense that the average mechanical pressure is $\bar{P}=\int d^3 r\,\langle N|\hat{P}(\bfvec{r})|N\rangle /V$.\\
As pointed out in I \cite{Nucleon-stability-I}, the spatial components of the EMT are not conserved, due to the lack of  spatial translational invariance induced by the presence of the static  confining interaction. Explicitly, the four-divergence of the EMT satisfies:
\begin{eqnarray}
 \partial_\mu \hat{T}^{\mu k}(x) &=& -q^\dagger(x)\,\nabla_k W(x)\,q(x).\label{VIRIAL0}
\end{eqnarray}
A virial theorem \cite{Lorce2021} follows immediately,
\begin{equation}
 \int d^3r \,\langle\Psi|  \hat{T}^{i i}(\bfvec{r}) |\Psi\rangle   -\int d^3r \,\langle\Psi|q^\dagger(\bfvec{r})\, \bfvec{r}\cdot \bfvec{\nabla}_{\bfvec{r}} W_C(\bfvec{r})\,q(\bfvec{r}) |\Psi\rangle=0,
\end{equation}
which is valid for any normalized stationary state $|\Psi\rangle$ (we implicitly consider $t=0$, i.e., $x=(0,\bfvec{r})$ in this discussion). In this relation the potential $W_C$ can be viewed as an external potential acting on each of the constituent quarks that make up the nucleon core. However this confining interaction originates from the existence of a  three-branches string junction which is at the origin of the  very  existence of the nucleon. Accordingly we can consider that the pressure operator should  be modified to incorporate the  presence of the confining interaction:
\begin{equation}
\hat{P}(x)=\frac{1}{3} \, \hat{T}^{i i}(x) \quad\to\quad \frac{1}{3} \, \hat{T}^{i i}(x)\,-\frac{1}{3} \,\,q^\dagger(x)\, \bfvec{r}\cdot \bfvec{\nabla}_{\bfvec{r}} W_C(\bfvec{r})\,q(x).
\end{equation}
If we now introduce the following notations: 
\begin{eqnarray}
\mathcal{P}_C(x)&=& \,q^\dagger(x)\, \bfvec{r}\cdot \bfvec{\nabla}_{\bfvec{r}}W_C(\bfvec{r})\,q(x)\equiv \hat{P}_C(x)\\ 
H'_C &=&\int d^3r \,q^\dagger(\bfvec{r})\, \bfvec{r}\cdot \bfvec{\nabla}_{\bfvec{r}}W_C(\bfvec{r})\,q(\bfvec{r})\\
P_C &=&\langle N|H'_C|N\rangle,
\end{eqnarray}
then the, physically motivated, modified pressure  operator finally reads:
\begin{equation}
\hat{P}(x)=\left(\frac{1}{3}\mathcal{H}_{QK}\,-\,\frac{1}{3} \,\mathcal{P}_C(x)\,+\,\frac{1}{3}\mathcal{H}_{\pi Q}\,+\,\mathcal{H}_{\pi c} 
\,-\,\frac{1}{3}\,\mathcal{H}_{\pi K}\,-\,\mathcal{H}_{\pi M}\right)(x).
\end{equation}
%%%%%%%%%%%%%%%%%%%%%%%%%%%%%%%%%%%%%%%%%%%%%%%%%%%%%%%%%%%%%%%%%%%%%%
\subsection{The von Laue stability condition }
As derived in section III-B of I, the virial theorem for the (free or bound) nucleon , $\int d^3r\,\langle \hat{P}\rangle(\bfvec{r})=0$, takes the explicit form,
\begin{eqnarray}
&&\frac{1}{3}\, E_{QK}(\mathcal{S})\,-\, \frac{1}{3}\,P_C(\mathcal{S})\,+\,\frac{1}{3}\, E_{\pi Q}(\mathcal{S}) \,+\, 
E_{\pi c}(\mathcal{S}) \,-\,\frac{1}{3}\,E_{\pi K}(\mathcal{S}) \,-\,E_{\pi M}(\mathcal{S}) =0,
\label{VONLAUEMED}\end{eqnarray}
which is nothing than the venerable von Laue stability condition \cite{vonLaue}, whereas the total energy of the bound nucleon is:
\begin{eqnarray}
E_0(\mathcal{S})&=&E_{QK}(\mathcal{S})\,+\,E_{QM}(\mathcal{S})\,+ E_C(\mathcal{S})\,+\, E_{\pi Q}(\mathcal{S}) \,+\, 
E_{\pi c}(\mathcal{S}) \,+\,E_{\pi K}(\mathcal{S}) \,+E_{\pi M}(\mathcal{S}). 
\end{eqnarray}
The physical meaning is very transparent: the quark Fermi pressure, $E_{QK}(\mathcal{S})/3$, is balanced by the negative bag pressure, $- E_C(\mathcal{S})/3,$ and the negative pion pressure  which  in practice is very close to the pion cloud kinetic pressure $-P_{\pi}\sim - E_{\pi K}(\mathcal{S})$. 
\\
Eq. \eqref{VONLAUEMED} is an exact result, but this von Laue condition is not necessarily satisfied by the Fock-space amplitude $\Phi\left({\bf p}_1, {\bf p}_2, {\bf p}_3\right)$, solution of the HF Dirac equation, according to the discussion given in subsection V-B of the companion paper. Hence, since our main focus concerns the question of mechanical stability of the nucleon, this von Laue stability condition may also serve as an alternative to the variational principle used in non-relativistic quantum mechanics and can be used as a physically motivated prescription to fix the Fock-space amplitude. In practice the three-quark Fock-space amplitude  will always  depend on the the Gaussian single-quark Fock-space amplitude :
\begin{equation}
\Phi(\bfvec{p})= R_0(p)= (2\pi)^{\frac{3}{2}}\,\left(\frac{b^2}{\pi}\right)^{\frac{3}{4}}\,e^{-b^2\,p^2/2}. 
\label{TRIALWF}
\end{equation}
Throughout the remainder of this paper, the value of the size parameter $b$ will be fixed by requiring that the von Laue stability condition is satisfied and the determination of the ``best value" of the quark core size parameter $b$ is the central  issue of the paper.\\

The explicit expression of $E_{QK}(\mathcal{S}),\,E_{QM}(\mathcal{S}),\, E_C(\mathcal{S}),\,P_C(\mathcal{S}),\, E_{\pi Q}(\mathcal{S}),\,
E_{\pi c}(\mathcal{S}), \,E_{\pi K}(\mathcal{S}), \,E_{\pi M}(\mathcal{S})$, derived in I \cite{Nucleon-stability-I} but generalized to the bound nucleon case, are listed in appendix ~\ref{app:Paper_1_summary}.
%%%%%%%%%%%%%%%%%%%%%%%%%%%%%%%%%%%%%%%%%%%%%%%%%%%%%%%%%%%%%%%%%%%%%%%%%%%%%%%%%%%%%%%%%%%%%%%%%%%%%%%%%%%%%%%%%%%%%%%%%%%%%%%%%%%%%%%%%%%%%%%%%%%%%%%%%%%%%%%%%%%%%%%%%%%%%%%%%%%%%%%%%%%%%%%%%%%%%%%%%%%%%%%
\section{Static EMT tensor: Energy density and pressure distribution in the in-medium nucleon}\label{sec:STATIC_EMT_CCM}
%%%%%%%%%%%%%%%%%%%%%%%%%%%%%%%%%%%%%%%%%%%%%%%%%%%%%%%%%%%%%%%%%%%%%%%%%%%%%%
%%%%%%%%%%%%%%%%%%%%%%%%%%%%%%%%%%%%%%%%%%%%%%%%%%%%%%%%%%%%%%%%%%%%%%%%%%%%%
\subsection{The static EMT tensor}
%%%%%%%%%%%%%%%%%%%%%%%%%%%%%%%%%%%%%%%%%%%%%%%%%%%%%%%%%%%%%%%%%%%%%%%%%%%%%
The static EMT tensor $t^{\mu\nu}(\bfvec{r})$ is defined  from the matrix element evaluated in the Breit frame of the  EMT tensor according to \cite{Polyakovor} :
\begin{eqnarray}
t^{\mu\nu}(\bfvec{r})&=&\int\frac{d^3\Delta}{(2\pi)^3}\, e^{-i\,\bfvec{\Delta}\cdot\bfvec{r}} \,\left\langle\frac{\bfvec{\Delta}}{2}\left| \,\hat{T}^{\mu\nu}(0,\bfvec{0})\,\right|-\frac{\bfvec{\Delta}}{2}\right\rangle=\int\frac{V d^3\Delta}{(2\pi)^3}
\left\langle\frac{\bfvec{\Delta}_N}{2}\left|\, \hat{T}^{\mu\nu}(0,\bfvec{r})\,\right|-\frac{\bfvec{\Delta}_N}{2}\right\rangle .
\end{eqnarray}
To also  account for the effect of confining pressure, we also introduce a Wigner density  that may be viewed as a ``static confining pressure", defined as follows:
\begin{eqnarray}
p_C(\bfvec{r})&=&\int\frac{d^3\Delta}{(2\pi)^3}\, e^{-i\,\bfvec{\Delta}\cdot\bfvec{r}} \,\left\langle\frac{\bfvec{\Delta}}{2}\left| \,\hat{P}_C(0,\bfvec{0})\,\right|-\frac{\bfvec{\Delta}}{2}\right\rangle=\int\frac{V d^3\Delta}{(2\pi)^3}
\left\langle\frac{\bfvec{\Delta}_N}{2}\left|\, \hat{P}_C(0,\bfvec{r})\,\right|-\frac{\bfvec{\Delta}_N}{2}\right\rangle .
\end{eqnarray}
At variance with the definition given in \cite{Goeke,Polyakov,Lorce2019,Lorce2021}, where a covariant normalization is used, the momentum states here are normalized either as in non relativistic quantum mechanics, ($\left\langle \bfvec{P}' |\bfvec{P}\right\rangle=(2\pi)^3\delta^{(3)}(\bfvec{P}'-\bfvec{P})$, or  to one in a large box of volume $V$, ($\left\langle \bfvec{P}'_N |\bfvec{P}_N\right\rangle=\delta_{\bfvec{P}'_N,\bfvec{P}_N})$. In the following we will omit the reference to the arbitrary time $t=0$, since in the Breit frame the two nucleon states, $|-\bfvec{\Delta}/2\rangle$ and $|\bfvec{\Delta}/2\rangle$, have the same energy. Owing to translational invariance, the matrix element of EMT tensor component can be identically expressed as: 
\begin{equation}
\left\langle\frac{\bfvec{\Delta}}{2}\left| \,\hat{T}_Y(0,\bfvec{0})\,\right|-\frac{\bfvec{\Delta}}{2}\right\rangle  =\int\frac{d^3 y}{V}  \,e^{i\,\bfvec{\Delta}\cdot\bfvec{y}} \,\left\langle\frac{\bfvec{\Delta}}{2}\left| \,\hat{T}_Y(\bfvec{y})\,\right|-\frac{\bfvec{\Delta}}{2}\right\rangle .
\end{equation}
The various components of the EMT tensor are observable quantities in the sense that they can be interpreted as Sachs  form factors \cite{RGS} (expressed in the Breit frame in analogy with the electromagnetic form factors \cite{Xiong}). These FFs   are  obtainable from the generalized parton distributions (GPDs) \cite{Polyakov,Lorce2019}, which are experimentally accessible through several exclusive processes, such as deeply virtual Compton scattering \cite{DVCS} and meson production \cite{MESPROD}. It has been proposed \cite{Polyakov} that the components  $t^{00}$ and the component
$t^{ii}$ can be related to the genuine energy density and the genuine mechanical isotropic pressure distribution  inside the nucleon according to:
\begin{eqnarray}
\varepsilon(r)= t^{00}(\bfvec{r})\qquad &\hbox{with:}&\qquad  M_N=\int 4\pi r^2\,dr\,\varepsilon(r)\\
p(r)=\frac{1}{3}t^{ii}(\bfvec{r})-\frac{1}{3}p_C(\bfvec{r}) \qquad &\hbox{with  the von Laue condition: }&\qquad
\int 4\pi r^2\,dr\,p(r)=0.
\end{eqnarray}
The explicit decompositions of the energy density and the pressure distribution are: 
\begin{eqnarray}
 \varepsilon(r;\mathcal{S})&=&\varepsilon_{QK}(r;\mathcal{S})\,+\,\varepsilon_{QM}(r;\mathcal{S})\,+ \varepsilon_C(r;\mathcal{S})\,+\, \varepsilon_{\pi Q}(r;\mathcal{S}) \,+\, 
\varepsilon_{\pi c}(r;\mathcal{S}) \,+\,\varepsilon_{\pi K}(r;\mathcal{S}) \,+\varepsilon_{\pi M}(r;\mathcal{S})\label{EPS00}\\  
p(r;\mathcal{S})&=&\frac{1}{3}\varepsilon_{QK}(r;\mathcal{S})\,-\frac{1}{3} p_C(r;\mathcal{S})\,+\,\frac{1}{3} \varepsilon_{\pi Q}(r;\mathcal{S}) \,+\, 
\varepsilon_{\pi c}(r;\mathcal{S}) \,-\,\frac{1}{3}\varepsilon_{\pi K}(r;\mathcal{S}) \,-\varepsilon_{\pi M}(r;\mathcal{S}).\label{PRESS0}
\end{eqnarray}
In an analogous way, and as discussed and justified in the introduction, one can associate with each local quark density operator of the type, $\hat{T}_\Gamma(\bfvec{r})=\bar{q}(\bfvec{r})\,\Gamma \,q(\bfvec{r})$, a density, i.e., a spatial distribution,  in the Wigner sense \cite{Wigner} according to:
\begin{eqnarray}
  t_\Gamma (\bfvec{r})=
\int \frac{V d^3\Delta}{(2\pi)^3}\, \left\langle \bfvec{\Delta}_N/2\right|\hat{T}_\Gamma (\bfvec{r}) \left| -\bfvec{\Delta}_N/2\right\rangle=\int \frac{V d^3\Delta}{(2\pi)^3}\,e^{-i\bfvec{\Delta}\cdot\bfvec{r}} \left\langle \bfvec{\Delta}_N/2\right|\hat{T}_\Gamma (0) \left| -\bfvec{\Delta}_N/2\right\rangle.
\end{eqnarray}
The explicit expression of $\varepsilon_{QK}(r;\mathcal{S}),\,\varepsilon_{QM}(r;\mathcal{S}),\, \varepsilon_C(r;\mathcal{S}),\,p_C(r;\mathcal{S}),\, \varepsilon_{\pi Q}(r;\mathcal{S}),\,
\varepsilon_{\pi c}(r;\mathcal{S}), \,\varepsilon_{\pi K}(r;\mathcal{S}), \,\varepsilon_{\pi M}(r;\mathcal{S})$, derived in I \cite{Nucleon-stability-I} but generalized to the bound nucleon case, are listed in appendix A.
%%%%%%%%%%%%%%%%%%%%%%%%%%%%%%%%%%%%%%%%%%%%%%%%%%%%%%%%%%%%%%%%%%%%%%%%%%%%%%%%%%%%%%%%%%%%%%%%%%%%%%%%%%%%%%%%%%%%%%%%%%%%%%
\section{Localized in-medium nucleon state: static nucleonic bag }\label{sec:STATIC_LOCALIZED_N}
%%%%%%%%%%%%%%%%%%%%%%%%%%%%%%%%%%%%%%%%%%%%%%%%%%%%%%%%%%%%%%%%%%
\subsection{Nucleon mass}
Although the nucleon state, associated with a factorized wave Fock-space amplitude as a product of three quark orbitals (Eq \eqref{WFNOCM}), is not a physical state with well defined three-momentum, anyway we will as usual model the nucleon using such a localized state $|N: \bfvec{0}_X\rangle$ with inclusion of the pion cloud. In that respect a rough estimate of the bare nucleon mass can be obtained by subtracting the spurious CM motion according to:
\begin{equation}
M_N=\left[\sqrt{E_0^2 -\langle P^2\rangle}\right]_{\mathcal{S}=M_0} = \left[\sqrt{E_0^2 -\frac{9}{2 b^2}}\right]_{\mathcal{S}=M_0}.
\end{equation}
As in the companion paper I \cite{Nucleon-stability-I}, we use the  NJLSet1 set of parameters \cite{Universe2025}, $M_q=356.7\,MeV$, $M_\pi=140\,MeV$, $F_\pi=91.7\,MeV$. The confining potential is taken  from the  FCM-inspired framework (Eqs. \eqref{CONFPOT}, \eqref{PRESSPOT}) with $\sigma=0.18\,GeV^2$ and $T_g=0.283\,fm$ \cite{Universe2025}. The pionic size parameter was chosen to be $c_\pi=0.8$ (i.e.,  $\rho_\pi=0.30\,fm$,  see section V-C of I) in order to satisfy the von Laue condition for an isolated nucleon,
\begin{eqnarray}
&&\left[\frac{1}{3}\, E_{QK}\,-\, \frac{1}{3}\,P_C\,+\,\frac{1}{3}\, E_{\pi Q} \,+\, 
E_{\pi c} \,-\,\frac{1}{3}\,E_{\pi K} \,-\,E_{\pi M}\right]_{\mathcal{S}=M_0} =0,
\end{eqnarray}
 so as to obtain a nucleon mass that is not excessively large while maintaining a reasonably small rms of the quark core.\\\
 The results of the calculation, for various values of the nuclear scalar field, are provided in table ~\ref{tab:parameters}. It can be seen that the value of the parameter $b_{virial}(\mathcal{S})$, for which the von Laue condition is satisfied, 
 \begin{eqnarray}
&&\frac{1}{3}\, E_{QK}(\mathcal{S})\,-\, \frac{1}{3}\,P_C(\mathcal{S})\,+\,\frac{1}{3}\, E_{\pi Q}(\mathcal{S}) \,+\, 
E_{\pi c}(\mathcal{S}) \,-\,\frac{1}{3}\,E_{\pi K}(\mathcal{S}) \,-\,E_{\pi M}(\mathcal{S}) =0,
\end{eqnarray}
increases as the quark mass decreases, i.e., as the nuclear scalar field $s$ increases in magnitude corresponding to the progressive restoration of chiral symmetry. A rough estimate of the in-medium nucleon mass can be obtained by subtracting the spurious CM motion according to:
\begin{equation}
M_N(\mathcal{S})=\sqrt{E_0^2(\mathcal{S}) -\langle P^2\rangle} = \sqrt{E_0^2(\mathcal{S}) -\frac{9}{2 b^2(\mathcal{S})}}.
\end{equation}
The important quantities for nuclear physics applications are the two dimensionless response parameters, the scalar coupling constant and the scalar susceptibility:
\begin{equation}
g_S=\frac{M_0}{F_\pi}\left(\frac{\partial M_N(\mathcal{S})}{\partial\mathcal{S} }\right)_{\mathcal{S}=M_0},\qquad
C=\frac{M_0^2}{2 M_N}\left(\frac{\partial^2 M_N(\mathcal{S})}{\partial\mathcal{S}^2 }\right)_{\mathcal{S}=M_0}.
\end{equation}
To second order in the nuclear scalar field $s$, the modification of the nucleon mass is parametrized as : 
\begin{equation}
M_N(s)-M_N(s=0) = -g_S\,|s|\, +\,M_N\,C \,\frac{s^2}{F^2_\pi}\,+\, \mathcal{O}(s^3). 
\end{equation}
Using the calculated values of $M_N(s)$ for $|s|/F_\pi=0.15, 0.3$, one can extract a rough  estimate of the two response parameters:
$$g_S=4.42\,\qquad C=0.33.$$
\begin{table*}[t]
\tabcolsep=0.5cm
\def\arraystretch{1.5}
\caption{\label{tab:parameters}%
Total energy and approximate mass of the CCM nucleonic static bag in the presence of the pion cloud on top of the confining potential  for various values of the nuclear scalar field $s$ and the corresponding constituent quark mass. The string tension is $\sigma=0.18\,GeV$  and the pion size parameter is $\rho_\pi =0.30\,fm$ ($c_\pi=0.8$). $b_{virial}$ is the value of the size parameter for which the von Laue condition for  mechanical stability is satisfied.}
\begin{tabular}{ccccccc}
\hline
$-s/F_\pi$ & $M_q=\mathcal{S}\, (MeV)$ & $b_{virial}\, (fm)$ & $E_0\,(GeV)$ & $\sqrt{E_0^2 -\langle P^2\rangle}\, (GeV)$ & $g_A$ & $r_B (fm)$    \\
\hline
$0$ & $356.7$ & $0.345$ & $1.616$ & $1.069$ & $1.13$ & $0.495$\\
\hline
$0.15$ & $303.2$ & $0.358$ & $1.562$ & $1.036$ & $1.09$ & $0.521$ \\
\hline
$0.30$ & $249.6$ & $0.369$ & $1.522$ & $1.015$ & $1.03$ & $0.548$ \\
\hline
$0.60$ & $142.7$ & $0.387$ & $1.506$ & $1.048$ & $0.87$ & $0.608$ \\
\hline
\end{tabular}
\end{table*}
 
%%%%%%%%%%%%%%%%%%%%%%%%%%%%%%%%%%%%%%%%%%%%%%%%%%%%%%%%%
\subsection{Static EMT tensor: energy density and pressure}
\begin{figure}
\includegraphics[width=0.45\textwidth,angle=0]{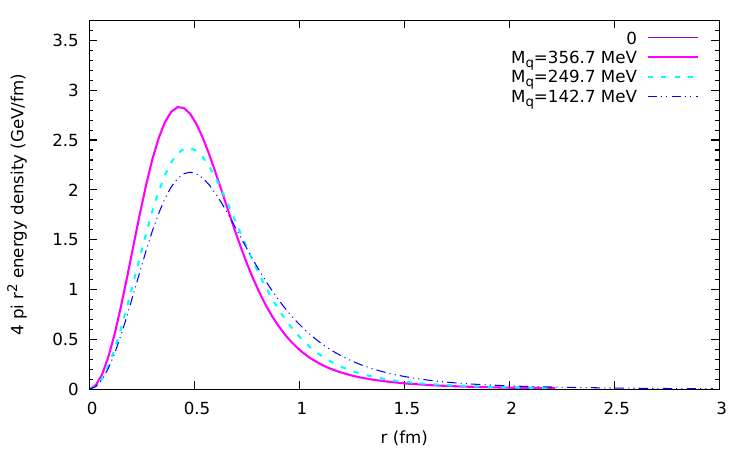}|
\includegraphics[width=0.45\textwidth,angle=0,height=5cm, keepaspectratio]{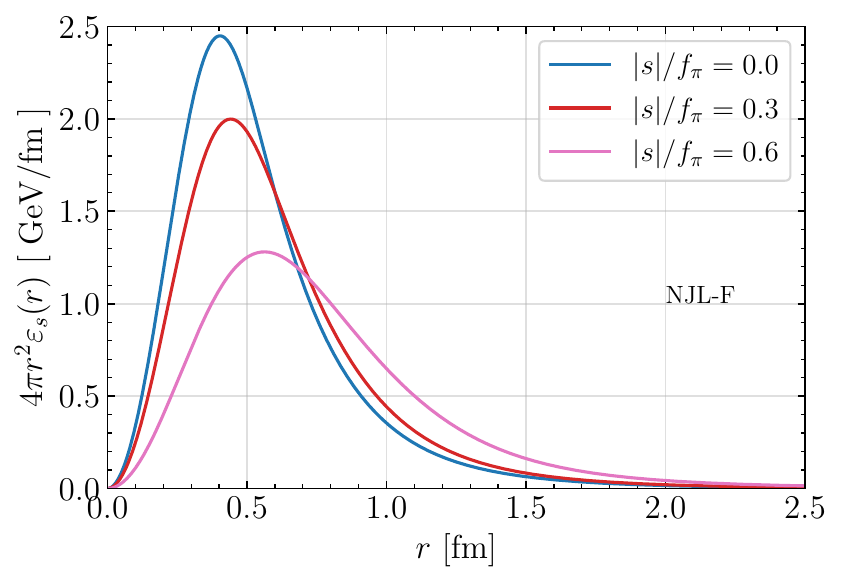}
\caption{Energy density distributions as a function of $r$ multiplied by $4\pi r^2$ for a  CCM nucleonic static  bag. Results are shown  for various values of the scalar field $|s|/F_\pi=0,0.3,0.6$ which correspond in the CCM model to constituent quark masses $M_q=356.7, 249.7,142.7\,MeV$. Left panel: static CCM nucleonic bag. Right panel: same curves but for  NJL-skyrmion model \cite{Pradhan}.}
\label{NOCMENERGY}
\end{figure}

\begin{figure}
\includegraphics[width=0.45\textwidth,angle=0]{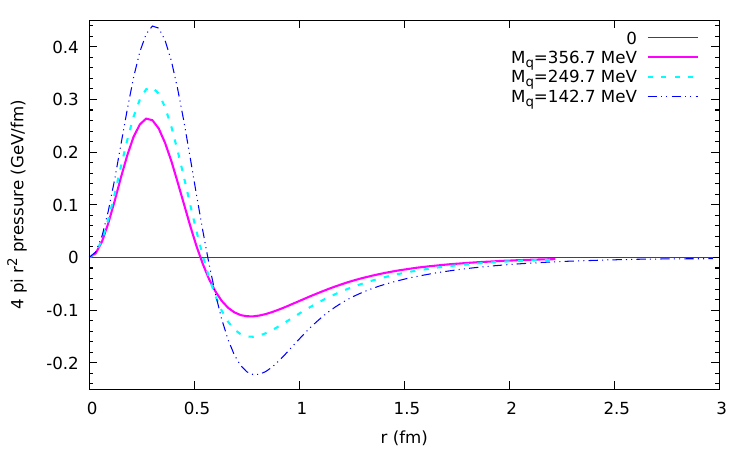}
\includegraphics[width=0.45\textwidth,angle=0,height=4.8cm, keepaspectratio]{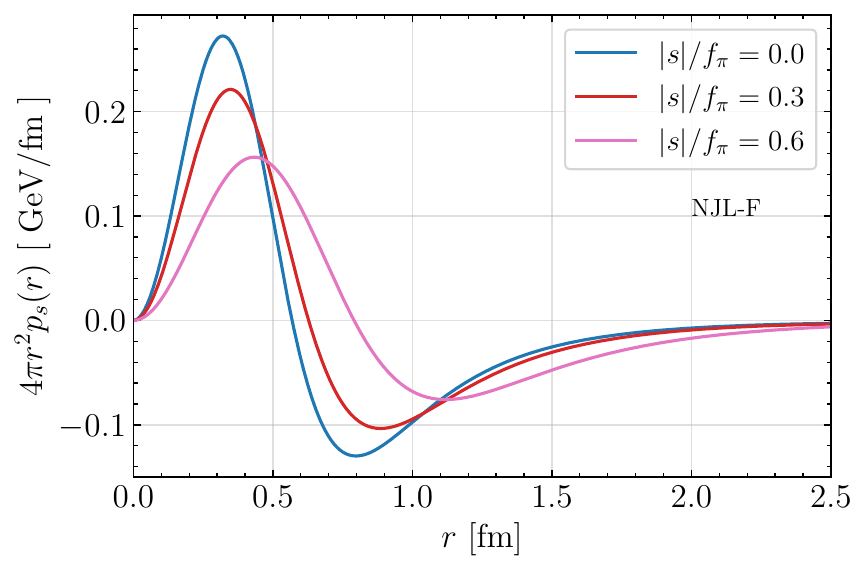}
\caption{Pressure distributions as a function of $r$ multiplied by $4\pi r^2$ for the CCM static nucleonic bag. Results are shown  for various values of the scalar field $|s|/F_\pi=0,0.3,0.6$ which correspond in the CCM model to constituent quark masses $M_q=356.7, 249.7,142.7\,MeV$. Left panel: static CCM nucleonic bag. Right panel: same curves but for  NJL-skyrmion model \cite{Pradhan}.} 
\label{NOCMPRESSURE}
\end{figure}

\begin{figure}
\includegraphics[width=0.45\textwidth,angle=0]{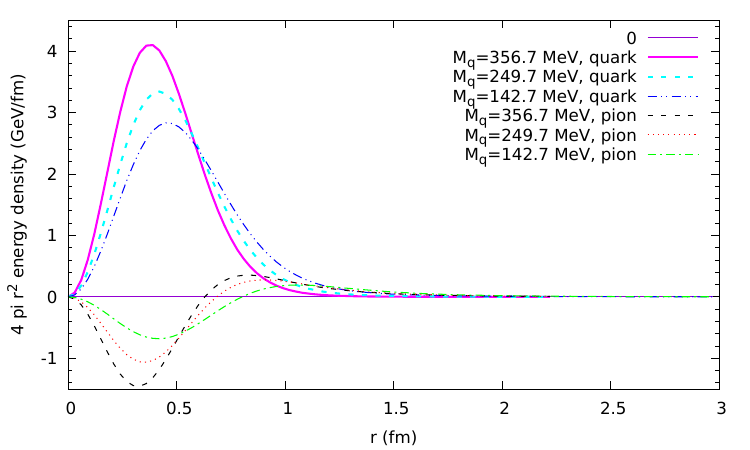}
\includegraphics[width=0.45\textwidth,angle=0]{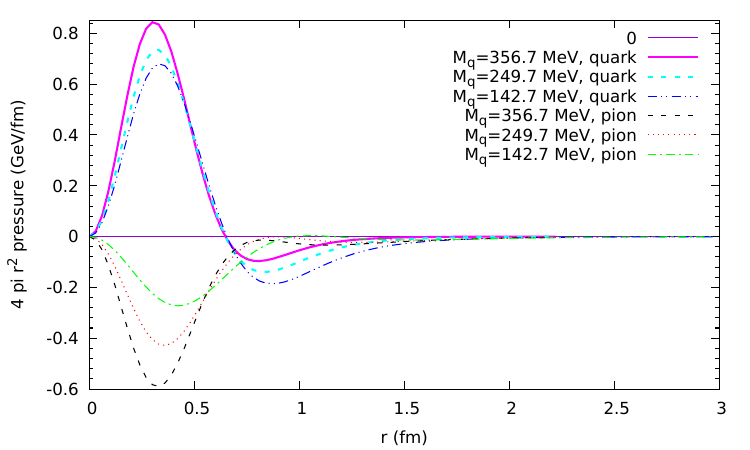}
\caption{The separate quark and pionic contributions to the energy density (left panel) and pressure (right panel) distributions  for   the static CCM nucleonic bag for various values of the scalar field $|s|/F_\pi=0,0.3,0.6$ which correspond  to constituent quark masses $M_q=356.7, 249.7,142.7\,MeV$ }
\label{NOCMDETAIL}
\end{figure}

%%%%%%%%%%%%%%%%%%%%%%%%%%%%%%%%%%%%%%%%%%%%%%%%%%%%%%%%%%%%%%%%%%%%%%%%%%%%%%%%%%%%%%%%%%%%%%%%%%%%%%%%%%%%%%%%%%%%%%%%%%%%%%
We  display the result of the calculation for the energy density distribution, $4\pi r^2\varepsilon(r)$, in Fig.~\ref{NOCMENERGY} (left panel). One can see that the energy density decreases overall and becomes more spread out as the quark mass decreases, reflecting the progressive restoration of chiral symmetry. It is striking to note that this result is very similar to the energy density obtained in a soliton/NJL-skyrmion model (Fig.~\ref{NOCMENERGY}, right panel) introduced in our recent paper \cite{Pradhan}, despite the fact that these two models are radically different: the chiral constraints are strong enough so that those two different chirally well-founded modelizations give the same phenomenology. As for the pressure (Fig.~\ref{NOCMPRESSURE}), we also observe very similar shapes for the isolated nucleon.  In contrast, for non-zero values of the nuclear scalar field a dramatically different behavior is observed. The magnitude of the short-distance pressure decreases in the NJL-skyrmion model, while it increases significantly in the CCM as the quark mass decreases. In fact, if one looks at the detailed contributions in Fig.~\ref{NOCMDETAIL} within the CCM, it becomes apparent that both the positive Fermi pressure from the quarks and the negative pion-induced pressure decrease individually. However, their difference increases, since the pion contribution decreases much more rapidly. This behavior is most likely due to the nonlocal coupling of the quarks to the pion field. It is as if the pion cloud progressively dilutes as a consequence of partial chiral
symmetry restoration.
\section{Momentum projected in-medium nucleon state: consequences for the nucleon matter equation of state }\label{sec:MOMENTUM_PROJECTED_N}
\subsection{Nucleon mass}
In the case of a projected momentum state as given  in Eq. \eqref{WFCM}, we explained  in I \cite{Nucleon-stability-I} that, achieving a satisfactory compromise between a sufficiently  low nucleon mass and a reasonably small quark-core size, required increasing the pionic size parameter to $c_\pi=1.2,\, (\rho_\pi=0. 37\,fm)$, and to modify the confining potential. We argued that it is physically plausible to replace the string tension $\sigma$ with an effective string tension $\sigma_E$ that rescales the size-dependent part of the confining interaction, as well as the associated confining pressure (the ``bag pressure”). If one sets its value  to $\sigma_E=0.12\,GeV^2$, the results of the calculation of the in-medium nucleon properties for various values of the nuclear scalar field, are tabulated in table~\ref{tab:Momentum_Projected_Nucleon}.%displayed on table II.
\begin{table*}[t]
\tabcolsep=0.3cm
\def\arraystretch{1.5}
\caption{\label{tab:Momentum_Projected_Nucleon}%
Mass of the nucleon, axial coupling constant and charge number radius for the nucleon momentum projected nucleon state for various values of the nuclear scalar field. The effective  string tension is $\sigma_E=0.12\,GeV^2$  and the pion size parameter is $\rho_\pi =0.37\,fm$ ($c_\pi=1.2$). $b_{E/virial}$ is the value of the size parameter for which the von Laue condition for mechanical stability is satisfied.}
\begin{tabular}{cccccccc}
\hline
$-s/F_\pi$ & $M_q=\mathcal{S}\, (MeV)$ & $b_{E/virial}\, (fm)$ & $E_Q\,(GeV)$ & $E_\pi\,(GeV)$ &$M_N(GeV)$ & $g_A$ & $r_B(fm)$   \\
\hline
$0$ & $356.7$ & $0.423$ & $1.371$ & $-0.315$ & $1.056$ & $1.21$ & $0.450$\\
\hline
$0.15$ & $303.2$ & $0.447$ & $1.248$ & $-0.270$ & $0.977$ & $1.17$ & $0.487$ \\
\hline
$0.30$ & $249.6$ & $0.474$ & $1.143$ & $-0.226$ & $0.917$ & $1.11$ & $0.529$ \\
\hline
$0.45$ & $196.2$ & $0.488$ & $1.063$ & $-0.183$ & $0.880$ & $1.05$ & $0.577$ \\
\hline
$0.60$ & $142.7$ & $0.521$ & $1.012$ & $-0.138$ & $0.874$ & $0.96$ & $0.634$ \\
\hline
\end{tabular}
\end{table*}

One very important conclusion is that the nucleon mass remains relatively stable, even possibly increases when the nuclear scalar field increases. This is due to the fact that the pionic attraction decreases more rapidly than the energy carried by the quarks. It can also be concluded that the nucleon survives,  that is,  the concept of in-medium nucleon remains valid, up to fairly high densities, but its internal structure evolves, as the gradual `evaporation' of the pion cloud gives way to a state resembling a pure quark bag (see the discussion about Fig.\ref{NOCMDETAIL}). As a consequence the effective mass of the in-medium nucleon, $M_N(s)\equiv\,M^*_N(s)$, exhibits a minimum at $|s|/F_\pi\sim 0.55$, which sets an upper bound beyond which the equation of motion for the scalar field, $V'_\chi(s)=-(\partial M^*_N/\partial s)\,\rho_S$, no longer has a conventional nucleonic  solution.\\
For density below or around normal nuclear matter density, i.e., around saturation density, the value of the scalar field is within the range $|s|/F_\pi< 0.4$. In this density domain, as in most of our previous phenomenological studies, the in-medium modification of the nucleon mass is parametrized as: 
\begin{eqnarray}
M_N(s)-M_N(s=0) &=& g_S\,s\, +\,M_N\,C \,\frac{s^2}{F^2_\pi}\,+\,\mathcal{O}(s^3)
\equiv-\alpha_N\,\frac{|s|}{F_\pi}\,+\beta_N\,\frac{s^2}{F^2_\pi}.
\end{eqnarray}
One can adjust the two parameters $\alpha_N, \beta_N$ to reproduce  the values of $M_N(s)$ for $|s|/F_\pi=0.15, 0.3$, with the result: 
\begin{equation}
 \alpha_N=587\,MeV,\qquad   \beta_N\,= 404\,MeV.
\end{equation}
On then obtain an estimate of the response parameters:
\begin{equation}
 g_S=6.39\,\qquad C=0.43.
\end{equation} 
As a major outcome of this  study, which is based on the mechanical stability of the nucleon,  we obtain results consistent with the values of the response parameters,  $g_S$ and $C$ (or $\kappa_{NS}$) that have been used in many of our previous phenomenological studies. In particular the values  of the scalar coupling constant  $g_S=6.52$ and the dimensionless scalar susceptibility $C=0.42$,  used in Ref.~\cite{Chanfray2024}  where a Hartree-Fock calculation, supplemented by pion-loop corrections, was carried out (see Fig.~5 of that article), fall definitely within the range of our results.
%%%%%%%%%%%%%%%%%%%%%%%%%%%%%%%%%%%%%%%%%%%%%%%%%%%%%%
\subsection{Static EMT tensor: energy density and pressure}
\begin{figure}
\includegraphics[width=0.45\textwidth,angle=0]{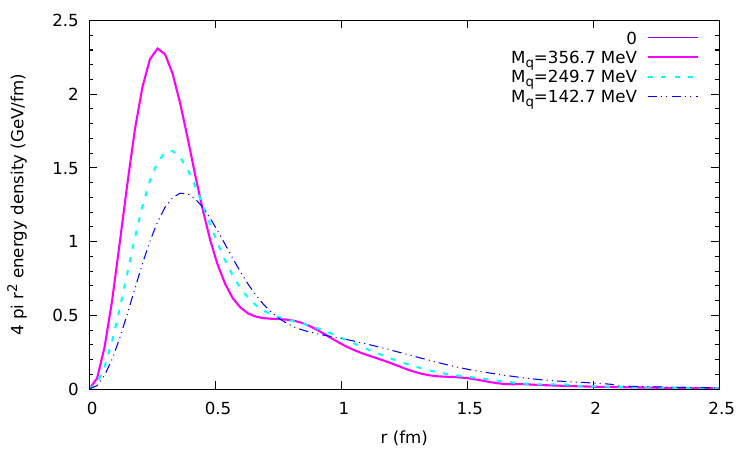}|
\includegraphics[width=0.45\textwidth,angle=0]{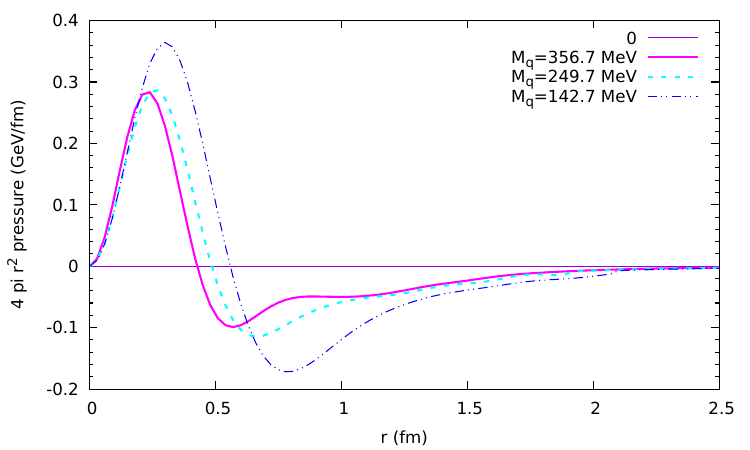}|
\caption{ Nucleon energy density (left panel) and pressure (right panel) distributions as a function of $r$ multiplied by $4\pi r^2$. Results are shown  for various of the scalar field $s/F_\pi=0,0.3,0.6$ which correspond  to constituent quark masses $M_q=356.7, 249.7,142.7\,MeV$. The reader may notice a slight oscillatory behavior at large values of $r$ which is most likely an artifact of the delicate double numerical integration of the product of two Bessel functions , which may become artificially divergent at large distance (see the discussion at the end of subsection VI-D of the companion paper \cite{Nucleon-stability-I}). 
In the present work we have chosen a large-distance cutoff such that the sum rule, $\int d^3r\,\rho_{B,S,K}(r)=1$, concerning the densities whose definition is recalled in Eqs. \eqref{EQKW}, \eqref{EQMW}, \eqref{EQCW}, \eqref{EQRHOB},  is satisfied to better than one part in a thousand for each of the three densities.}
\label{CMENERGY}
\end{figure}

\begin{figure}
\includegraphics[width=0.45\textwidth,angle=0]{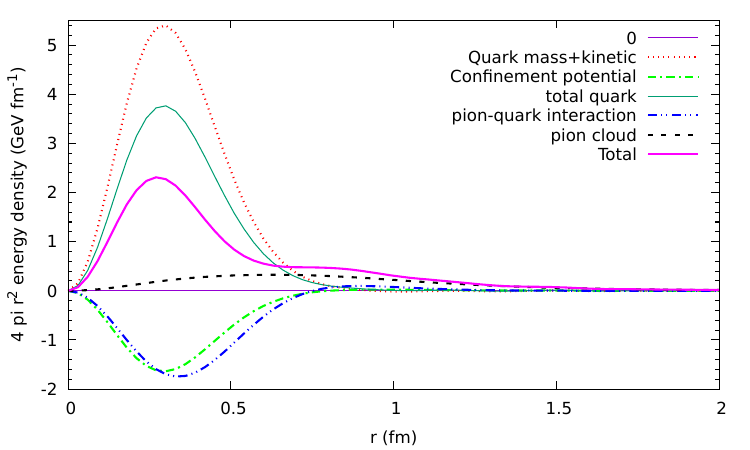}|
\includegraphics[width=0.45\textwidth,angle=0]{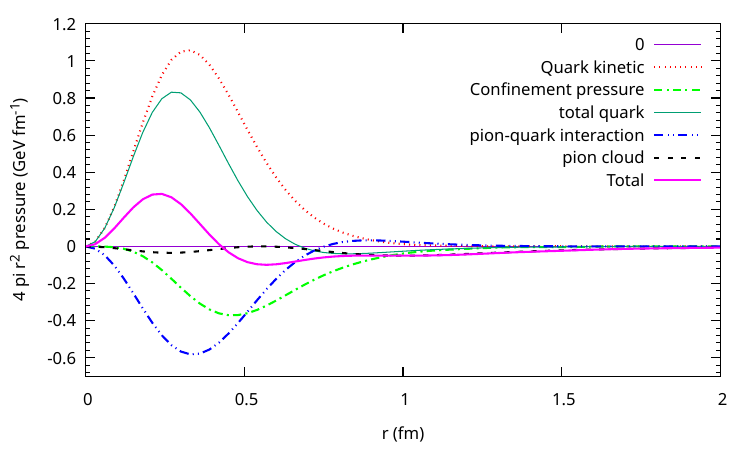}|
\caption{The various contributions to the energy density (left panel, 6a) and to the pressure   distribution (right panel, 6b) for $s=0$, i.e., for an isolated nucleon. } 
\label{QUARKDENSITY}
\end{figure}

%%%%%%%%%%%%%%%%%%%%%%%%%%%%%%%%%%%%%%%%%%%%%%%%%%%%%%%%%%%%%%%%%%%%%%%%%%%%%%%%%%%%%%%%%%%%%%%%%%%%%%%%%%%%%%%%%%%%%%%%%%%%
We  display the result of the calculation for the energy distribution in Fig.~\ref{CMENERGY} (left panel). One can see again that the energy density decreases overall and becomes more spread out as the quark mass decreases, reflecting the progressive restoration of chiral symmetry. One can also observe, in the case of the nucleon in vacuum, a sort of bump around $r=0.7\,fm$. As shown in the left panel of Fig.~\ref{QUARKDENSITY},  this is related to the fact that the pion cloud and the valence quarks occupy two seemingly separate regions, with the pion cloud extending well beyond the quark core. Regarding the pressure distribution (see Fig~\ref{CMENERGY}, right panel), we again observe a significant increase in the central region. This effect can be qualitatively considered as associated with the ``evaporation" of the pion cloud, which strongly reduces the negative pion pressure in that region. We also observe in the right panel of Fig.~\ref{QUARKDENSITY} that the positive quark Fermi pressure in an isolated nucleon is equilibrated with similar weight by both the negative pion pressure and the negative confinement pressure. However, with increasing density, i.e., larger magnitude of the nuclear scalar field, due to the progressive dilution of the pion field, the equilibration of the Fermi pressure has to be mainly ensured by a growing confining pressure. This necessitates an increase of the size parameter $b_E$ and consequently a spreading of the quark core, which can be seen as a consequence of partial chiral symmetry restoration.\\
It is also interesting to compare our results for the nucleon in vacuum with existing calculations. This allows us to verify that the pressure distribution is somehow different from that obtained with the chiral quark-soliton model presented in Fig. 4 of the Ref. \cite{Goeke}, where the magnitude of the pressure in the quark core is twice smaller than in our case. As for the energy distribution presented in Fig. 1 of Ref. \cite{Goeke}, it is quite different in that it does not show, as in our case, two clearly separated regions associated with the quarks and the pion cloud. In fact, the result from that reference more closely resembles our results obtained with a lower constituent quark mass. Our result for the pressure distribution has a common feature with the one presented in Fig. 7b of Ref. \cite{Lorce2019}, namely that the pressure in the central region is significantly larger than in the chiral quark model case. In this latter paper, the EMT components  are obtained directly from the generalized parton distributions (GPDs), which are accessible in several exclusive processes, such as deeply virtual Compton scattering \cite{DVCS} and meson production \cite{MESPROD}, via a modeling of the gravitational form factors. An important point is that the energy distribution shown in the right panel of Fig.~\ref{NOCMENERGY} of this paper has a significantly larger extent than both the results from Ref. \cite{Goeke} and our own, despite the relatively long tail associated with the pion cloud in our approach.

%%%%%%%%%%%%%%%%%%%%%%%%%%%%%%%%%%%%%%%%%%%%%%%%%%%%%%%%%%%%%%%%%%%%%%%%%%%%%%%%%%%%%%%%%%%%%%%%%%%%%%%%%%%%%%%%%%%%
%%%%%%%%%%%%%%%%%%%%%%%%%%%%%%%%%%%%%%%%%%%%%%%%%%%%%%%%%%%%%%%%%%%%%%%%%%%%%%%%%%%%%%%%%%%%%%%%%%%%%%%%%%%%%%%%%%
\section{perspectives: consequences for  deep interior neutron star equation of state }\label{sec:NUCLEON_NS}
\begin{figure}
\includegraphics[width=0.45\textwidth,angle=0]{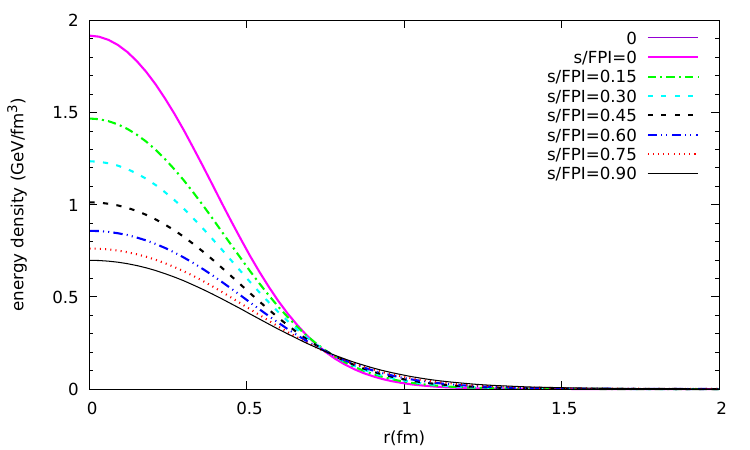}|
\includegraphics[width=0.45\textwidth,angle=0]{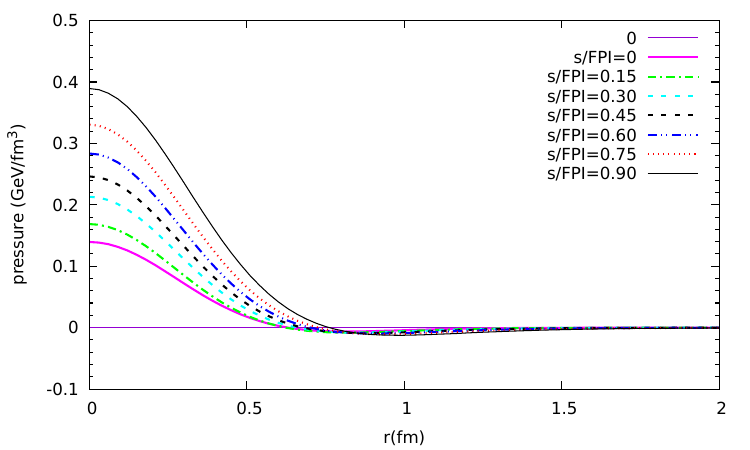}|
\caption{The energy density (left panel) and the pressure (right panel) distribution in the the cloudy quark core (QC) for various values of $|s|/F_\pi$. }
\label{HARDENERGY}
\end{figure}

\begin{figure}
\includegraphics[width=0.8\textwidth,angle=0]{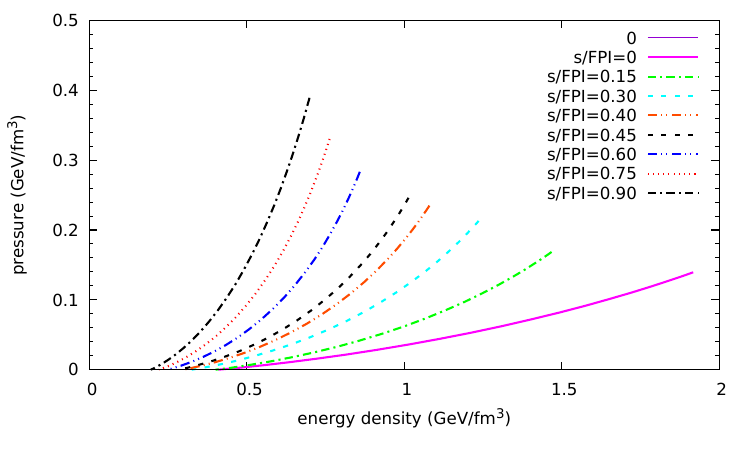}|
\caption{The EoS of the hard deconfined quark matter identified with the EoS of the interior of the QC's, for increasing values of the scalar field corresponding to decreasing values of the constituent quark mass as given in table~\ref{tab:parameters}.  }
\label{HARDEOS}
\end{figure}

Let us now consider the possible consequences for dense matter,  as realized in the deep interior of neutron stars. It has been conjectured that at sufficiently high baryon density, the equation of state (EoS) of bulk matter can be identified with the EoS of the nucleon core, establishing a direct link between nucleon structure and the EoS of the de-confined matter \cite{Lorce2019,Fuku,Pradhan}. As proposed in  \cite{Pradhan}, one can imagine the so-called hard-deconfined matter  \cite{Fuku} as a juxtaposition of quark cores. Here, we will assume that these quark cores (QC) and their attached pion cloud are not physical nucleon states with well-defined momentum but can be identified with the static nucleonic bag with a frozen CM position, namely the states $|N: \bfvec{X}_N\rangle$.  Hence, we reconsider these particular states, but with the parameters $c_\pi=1.2$ and $\sigma=0.12\, GeV^2$, for various values of the nuclear scalar field $s$ or equivalently the in-medium constituent quark mass, $M_q=M_0 (F_\pi-s)/F_\pi$. The properties of these ``cloudy"  QC's are given in table~\ref{tab:Energy_QC_Properties}. To third order in the nuclear scalar field $s$, the $s$-dependence of the cloudy QC energy can be parametrized, as in the case  of the nucleon mass,  as: 
\begin{eqnarray}
E_0(s)=E_0(s=0) -\alpha_C\,\frac{|s|}{F_\pi}\,+\beta_C\,\frac{s^2}{F^2_\pi}\,-\,\gamma_C\,\frac{|s|^3}{F^3_\pi}.
\end{eqnarray}
One can adjust the three parameters, $\alpha_C, \beta_C, \gamma_C$, to reproduce  the values of $E_0(s)$ for $|s|/F_\pi=0.15, 0.3, 0.45$, yielding the following values:
\begin{equation}
 \alpha_C=530\,MeV,\qquad   \beta_C\,= 266\,MeV,\qquad \gamma_C\,= -149\, MeV.
\end{equation}
\begin{table*}[t]
\tabcolsep=0.3cm
\def\arraystretch{1.5}
\caption{\label{tab:Energy_QC_Properties}%
Energy and quark core charge radius of the nucleonic cloudy quark core (QC) for various values of the nuclear scalar field. The effective  string tension is $\sigma=0.12\,GeV^2$  and the pion size parameter is $\rho_\pi =0.37\,fm$ ($c_\pi=1.2$). $b_{virial}$ is the value of the size parameter $b$ for which the von Laue condition for mechanical stability is satisfied.}
\begin{tabular}{cccccccc}
\hline
$-s/F_\pi$ & $M_q=\mathcal{S}\, (MeV)$ & $b_{virial}\, (fm)$ & $E_Q\,(GeV)$ & $E_\pi\,(GeV)$ &$E_0(GeV)$ &  $r_B(fm)$   \\
\hline
$0$ & $356.7$ & $0.425$ & $1.559$ & $-0.189$ & $1.370$ &  $0.595$\\
\hline
$0.15$ & $303.2$ & $0.440$ & $1.464$ & $-0.167$ & $1.297$ &  $0.625$ \\
\hline
$0.30$ & $249.6$ & $0.453$ & $1.384$ & $-0.145$ & $1.239$ &  $0.656$\\
\hline
$0.45$ & $196.2$ & $0.466$ & $1.323$ & $-0.124$ & $1.199$ &  $0.690$\\
\hline
$0.60$ & $142.7$ & $0.477$ & $1.283$ & $-0.101$ & $1.182$ &  $0.731$ \\
\hline
$0.75$ & $89.2$ & $0.487$ & $1.272$ & $-0.071$ & $1.201$ &  $0.776$ \\
\hline
$0.90$ & $35.7$ & $0.502$ & $1.293$ & $-0.024$ & $1.269$ &  $0.839$ \\
\hline
\end{tabular}
\end{table*}
The first observation is that these ``cloudy" quark-core bags are significantly larger than the physical nucleon states. Again, as the scalar field strength increases, which corresponds to more restored chiral symmetry, the quarks become progressively delocalized. This behavior is also reflected in the energy density profiles, as illustrated in~Fig.\ref{HARDENERGY} (left panel). The corresponding pressure distributions are shown in the right panel. The EoS for various values of the scalar field is displayed in Fig~\ref{HARDEOS}.\\
To construct a density-dependent EoS, we proceed as follows. The number of QCs per unit volume is given by the bulk baryonic density $\rho_N$, and the bulk energy density is: 
\begin{equation}
  \varepsilon_N= E_0(s)\,\rho_N(s).  
\end{equation}
The correspondence between the scalar field and the baryonic density is determined  by the equation of motion of the scalar field. Since the QC's are frozen, their scalar density can be replaced by  the baryonic density:
\begin{equation}
 V'_\chi(s)=-\frac{\partial E_0(s)}{\partial s}   \,\rho_N.
\end{equation}
If one uses the  NJL chiral potential from equation \eqref{vchiNJL}, the bulk baryonic density and the bulk energy density can be directly obtained as a function of the scalar field, according to:
\begin{eqnarray}
 \rho_N (s) &=&\rho_c\,\frac{|s|}{F_\pi}\,\frac{1\,-\,\frac{3}{2}\,\frac{|s|}{F_\pi}(1\,-\,C_\chi)}{1\,-2\,\frac{\beta_C}{\alpha_C}\,\frac{|s|}{F_\pi}\,-\,3\frac{|\gamma_C|}{\alpha_C}\,\frac{s^2}{F^2_\pi}} \qquad\hbox{with}\qquad \rho_c= \frac{M^2_\sigma\,F^2_\pi}{\alpha_C}\simeq 1.06\,fm^{-3},\\
 \varepsilon_N (s) &= &E_0(s)\,\rho_c\,\frac{|s|}{F_\pi}\,\frac{1\,-\,\frac{3}{2}\,\frac{|s|}{F_\pi}(1\,-\,C_\chi)}{1\,-2\,\frac{\beta_C}{\alpha_C}\,\frac{|s|}{F_\pi}\,-\,3\frac{|\gamma_C|}{\alpha_C}\,\frac{s^2}{F^2_\pi}},
\end{eqnarray}
where $M_\sigma=716\,MeV$ is the scalar meson mass in the NJLset1 parameter-set, and $C_\chi=0.488$. 
While the energy density inside each QC has a spatial profile, we implicitly assume that, on average, the quarks experience an effective energy density equal to $\varepsilon_{N}$. Thus, at a given baryon density $\rho_N$, with the corresponding nuclear scalar field $s(\rho_N)$, we determine the radial distance $r=r_{N}$ from the center of the QC obtained for this given $s(\rho_N)$ at which the local energy density $\varepsilon(r_N)$ coincides with the bulk density $\epsilon_N$. To this $\varepsilon(r_N)$, we associate, from the QC EOS, the pressure  $p(r_N)$, which we identify with the bulk pressure. In this way, we map the QC energy-density profile to the bulk average density to obtain the EoS. This analysis reveals that, as in our previous study \cite{Pradhan}, $\varepsilon_N$ increases with $\rho_N$, consistent with a denser packing of QCs. However, the maximum local energy density within the cloudy quark core, $\varepsilon(r)$, decreases as the in-medium scalar field $s$ diminishes with increasing density, signaling partial chiral restoration. This competition naturally defines a breakdown density $\rho_N$ (or equivalently a limiting $\varepsilon_{N}$), beyond which the cloudy quark core  EoS ceases. Specifically, the density-dependent EoS truncates at critical densities such that $\varepsilon (r=0)=\varepsilon_N\simeq 1\, GeV\cdot fm^{-3}$ and the maximum reachable pressure is $p\simeq 0.25\, GeV\cdot fm^{-3}$. On Fig.~\ref{HARDEOS}, this corresponds to the endpoint of the EoS corresponding to $|s|/F_\pi=0.45$. The corresponding density is approximately $\rho_N=0.8\,fm^{-3}$, which is significantly smaller than the breakdown density obtained with the NJL-soliton model of Ref. \cite{Pradhan}. We postpone the explicit construction of this EOS for future publication, using the exact baryonic and quark energy density profiles and the NJL chiral potential.\\

While the knowledge of the equation of state and the pressure distribution inside the nucleon is  itself a very interesting topic that motivates an important experimental effort (see, e.g., Ref \cite{DVCS}), the difficult question is how to make the connection between this internal mechanical pressure and the thermodynamic pressure of bulk matter. The validity of the mapping between nucleon structure properties and the equation of state of dense uniform matter that we propose in this paper has only an exploratory character, since it is certainly plagued by important finite-size (i.e., surface) effects, limiting its applicability to the region near the center of the quark core which corresponds  to the highest baryonic densities. To explicitly test the limitations of  this approach deserves  a future work for a comparison of the mechanical pressure with the thermodynamic pressure  $p_{th}(r)$,  defined using the local properties of the nucleon as: 
\begin{equation}
p_{th}(r)=\rho_B(r)\frac{\partial\varepsilon(r)}{\partial\rho_b(r)}-\varepsilon(r).
\end{equation}

\section{Conclusion}\label{sec:CONCLUSION}
%This article, together with the preceding one \cite{Nucleon-stability-I}, aimed to 

In this article, along with the preceding one~\cite{Nucleon-stability-I}, we determine in a self-consistent way the evolution of the properties of the composite nucleon when it is placed in a nuclear scalar field. Compared with our previous review paper \cite{Universe2025}, a major improvement has been proposed through the use of the von Laue mechanical stability condition to fix the size of the nucleon (i.e., the range of the constituent quark Fock-space amplitude), as well as a precise formulation of a translationally invariant nucleon wave function leading to a well-defined nucleonic three-momentum.
This also makes it possible to define, in the Wigner sense, local quark densities (the quark kinetic energy density operator,  the scalar density operator, or the quark number density operator) and to treat consistently the string-like confining interaction that explicitly breaks translational invariance. We also propose—for the first time, to our knowledge, in this type of model—a detailed formalism for defining and evaluating the local energy density and the pressure distribution (more generally, the static energy–momentum tensor) inside the nucleon, making it possible to visualize the delicate balance between the Fermi pressure and the negative pressure arising from the pion cloud and the confining interaction. \\

As a major outcome, we obtain values of the response parameters, $g_S=6.39$ and $C=0.43$, governing the binding energy and the saturation of nuclear matter, consistent with those used in many of our previous phenomenological studies \cite{Chanfray2024}. Another important achievement is that, as the density increases, one observes a gradual increase in the size of the nucleon and a progressive `dilution' or `evaporation' of the pion cloud, a clear manifestation of the progressive restoration of chiral symmetry. This then translates into a stiffening of the nucleon's internal equation of state (energy density vs. pressure), which could manifest as a so-called hard deconfined matter phase, intermediate between ordinary nuclear matter and pure quark matter in massive neutron stars.
\section{Acknowledgment}
The authors acknowledge the support of the CNRS-IN2P3 MAC masterproject, the project RELANSE ANR-23-CE31-0027-01 of the French National Research Agency (ANR), the European Union’s Horizon 2020 research and innovation program under grant agreement STRONG–2020-No824093. The authors also acknowledge Jérome Margueron for valuable discussions and his continued interest in this work. One of the authors, GC, further acknowledges C. Lorcé and P. Schweitzer for highly informative email exchanges regarding the virial theorem and the static energy-momentum tensor. 
%%%%%%%%%%%%%%%%%%%%%%%%%%%%%%%%%%%%%%%%%%%%%%%%%%%%%%%%%%%%%%%%%%%%%%%%%%%%%%%%%%%%%%%%%%%%%%%%%%%%%%%%%%%%%%%%%%%%%%%%%%%%%%%%%%%%%%%%%%%%%%%%%%%%%%%%%%%%%%%%%%%%%%%%%%%%%%%%%%%%%%%%%%%%%%%%%%%%%%%%%%%%%%%%%%%%%%%%%%%%%%%%%%%%%%%%%%%%ùù
\appendix 
	
\section{Summary of results derived in I }\label{app:Paper_1_summary}

\subsection{Quark energy}
For the localized CCM static bag, $|N: \bfvec{0}_X\rangle\,\equiv\,|N(\bfvec{X}_N=0)\rangle$, the various contributions to the quark energy and the average pressure, are  given by:
\begin{eqnarray}
 E_Q &=& E_{QK}+ E_{QM}+E_C= \int d^3 r\,\varepsilon_Q(r)= \int d^3 r\,\varepsilon_{QK}(r)\,+\,\int d^3 r\,\varepsilon_{QM}(r)\,+\,\int d^3 r\,\varepsilon_C(r)  \nonumber\\
 P_C&=& -\int d^3 r\,\mathcal{P}_C(r)\\
 \varepsilon_{QK}(r)&=&\left\langle N:\bfvec{0}_X\right|q^\dagger(\bfvec{r})\left[-i\bfvec{\alpha}\cdot\vec\nabla_{\bfvec{r}} \right]\,q(\bfvec{r}) \left|N:\bfvec{0}_X\right\rangle\nonumber\\ 
 &=&3\,\left( u(r)v^{\prime}(r)-u^{\prime}(r) v(r)\,+\,\frac{2 u(r) v(r)}{r}\right)\\
 \varepsilon_{QM}(r)&=&\left\langle N:\bfvec{0}_X\right|q^\dagger(\bfvec{r}) \beta \mathcal{S}\,q(\bfvec{r}) \left|N:\bfvec{0}_X\right\rangle=3\,\mathcal{S}\,\left(u^2(r) - v^2(r)\right)\\
 \varepsilon_{C}(r)&=&\left\langle N:\bfvec{0}_X\right|q^\dagger(\bfvec{r})\,W_C(\bfvec{r})\,q(\bfvec{r}) \left|N:\bfvec{0}_X\right\rangle = 3\,W_C(r)\,\left(u^2(r) +v^2(r)\right)\\
 -\mathcal{P}_C(r)&=&-\left\langle N:\bfvec{0}_X\right|q^\dagger(\bfvec{r})\,\bfvec{\nabla}_{\bfvec{r}}W_C(\bfvec{r})\,q(\bfvec{r}) \left|N:\bfvec{0}_X\right\rangle=
 -3\,W'_C(r)\,\left(u^2(r) +v^2(r)\right).
\end{eqnarray}
Alternatively, the quark kinetic and mass energies can be calculated more directly in momentum space:
\begin{eqnarray}
E_{QK} &=& 3\int \frac{d^3p}{(2\pi)^3}\,\frac{p^2}{E_p}\,\Phi^2(p)\\
E_{QM} &=& 3\int \frac{d^3p}{(2\pi)^3}\,\frac{\mathcal{S}^2}{E_p}\,\Phi^2(p)\\
E_{QK}+E_{QM}&=&3\int \frac{d^3p}{(2\pi)^3}\,E_p\,\Phi^2(p).
\end{eqnarray}
For the physical nucleon state, $|N: \bfvec{0}_P\rangle\,\equiv\,|N(\bfvec{P}_N=0)\rangle$ one has formally  the same expressions for the kinetic energy and the mass energy but with the $b$ parameter replace by $b_E=\sqrt{3/2}\,b$. For reasons  explained in section I of I, the energy and the average pressure associated  with the confining interaction necessitates the introduction of the baryonic Wigner density:
\begin{eqnarray}
E_C&=& 3\int d^3r\,W_C(r)\rho_B(r),\qquad  -P_C=- 3\int d^3r\,W'_C(r)\rho_B(r),\nonumber\\
\rho_B(r)&=&3\,\int\frac{d^3\Delta\,d^3x}{(2\pi)^3}\, e^{5 b^2\Delta^2/24}\,j_0(\Delta r)\,j_0(\Delta x)\,
\left(\psi^\dagger_E \psi_E\right)(x).
\end{eqnarray}
\subsection{Pionic energy}
For the localized CCM static bag, $|N: \bfvec{0}_X\rangle\,\equiv\,|N(\bfvec{X}_N=0)\rangle$,  the various contributions to the pionic energy are given by:
\begin{eqnarray}
E_{\pi M}&=&\frac{3}{2}\left(\frac{1}{2 F_\pi(\mathcal{S})}\right)^2 \int{d^3k\over 
(2\pi)^3} \frac{{\bf k}^2 \,M^2_\pi(\mathcal{S})}{2 \omega^2_k}\,\Gamma^2(k)\,\bigg[\frac{1}{\omega_k}\frac{1}{\omega_k+\epsilon_{N {\bf k}}}
+\frac{1}{(\omega_k+\epsilon_{N {\bf k}})^2}\nonumber\\
&&+\frac{32}{25}\,\bigg(\frac{1}{\omega_k}\frac{1}{\omega_k+\epsilon_{\Delta{\bf
k}}} + \frac{1}{(\omega_k+\epsilon_{\Delta{\bf k}})^2}\bigg)\bigg]\label{EPIMF}\\
E_{\pi K}&=&{3\over 2}\left({1\over 2 F_\pi(\mathcal{S})}\right)^2 \int{d^3k\over 
(2\pi)^3} \frac{{\bf k}^4}{2 \omega^2_k}\,\Gamma^2(k)\,\bigg[\frac{1}{\omega_k}\frac{1}{\omega_k+\epsilon_{N {\bf k}}}
+\frac{1}{(\omega_k+\epsilon_{N {\bf k}})^2}\nonumber\\
&&+\frac{32}{25}\,\bigg(\frac{1}{\omega_k}\frac{1}{\omega_k+\epsilon_{\Delta{\bf
k}}} + \frac{1}{(\omega_k+\epsilon_{\Delta{\bf k}})^2}\bigg)\bigg]\label{EPIKF}\\
E_{\pi c}&=&-{3\over 2}\left({1\over 2 F_\pi(\mathcal{S})}\right)^2 \int{d^3k\over 
(2\pi)^3} \frac{{\bf k}^2}{2}\,\Gamma^2(k)\,\bigg[\frac{1}{\omega_k}\frac{1}{\omega_k+\epsilon_{N {\bf k}}}
+\frac{1}{(\omega_k+\epsilon_{N {\bf k}})^2}\nonumber\\
&&+\frac{32}{25}\,\bigg(\frac{1}{\omega_k}\frac{1}{\omega_k+\epsilon_{\Delta{\bf
k}}} + \frac{1}{(\omega_k+\epsilon_{\Delta{\bf k}})^2}\bigg)\bigg]\nonumber\\
&&+\frac{3}{2}\left({1\over 2 F_\pi(\mathcal{S})}\right)^2 \int{d^3k\over 
(2\pi)^3}\,k^2\,\Gamma^2(k)\,\bigg[\frac{1}{\omega_k}\frac{1}{\omega_k+\epsilon_{N {\bf k}}} +\frac{32}{25}\,\frac{1}{\omega_k}\frac{1}{\omega_k+\epsilon_{\Delta{\bf
k}}} \bigg]\label{EPICF}\\
E_{\pi Q}&=&-3\left({1\over 2 F_\pi(\mathcal{S})}\right)^2 \int{d^3k\over 
(2\pi)^3} \,k^2\,\Gamma^2(k)\,\bigg[\frac{1}{\omega_k}\frac{1}{\omega_k+\epsilon_{N {\bf k}}} +\frac{32}{25}\,\frac{1}{\omega_k}\frac{1}{\omega_k+\epsilon_{\Delta{\bf
k}}} \bigg].\label{EPIQF}
\end{eqnarray}
For the physical nucleon state, $|N: \bfvec{0}_P\rangle\,\equiv\,|N(\bfvec{P}_N=0)\rangle$, one has formally  the same expressions for the various contributions to the pionic energy but with the $b$ parameter replaced by $b_E=\sqrt{3/2}\,b$, and the form factor $\Gamma(k)$ replaced by $\Gamma(k)\exp(5 b^2 k^2/24)$, namely:
\begin{equation}
 \Gamma(k)=  e^{5 b^2k^2/24}\,P(k)\,\frac{5}{3}\int d^3x\,\left[j_0(kx)\left(u_E^2(x)-\frac{v_E^2(x)}{3}\right)-\frac{4}{3}\,j_2(kx) \,v_E^2(x)\right].   
\end{equation}
%%%%%%%%%%%%%%%%%%%%%%%%%%%%%%%%%%%%%%%%%%%%%%%%%%%%%%%%%%%%%%%%%%%%%%%%%%%%%%%%%%%%%%%%%%%%%%%%%%%%%%%%%%
\subsection{Quark core contribution to the energy density and pressure}
%%%%%%%%%%%%%%%%%%%%%%%%%%%%%%%%%%%%%%%%%%%%%%%%%%%%%%%%%%%%%%%%%%%%%%%%%%%%%%%%%%%%%%%%%%%%%%%%%%%%%%%%%%%%%%%%%%%
For the localized CCM static bag, $|N: \bfvec{0}_X\rangle\,\equiv\,|N(\bfvec{X}_N=0)\rangle$, the various contributions to the quark energy densities and pressure distributions, are  given by:
\begin{eqnarray}
 \varepsilon_{QK}(r)/3&=&\left(\psi\right)^\dagger(\bfvec{r})[-i\bfvec{\alpha}\cdot\vec\nabla_{\bfvec{r}} ]\,\left(\psi\right)(\bfvec{r}) 
 = u(r)v^{\prime}(r)-u^{\prime}(r) v(r)\,+\,\frac{2 u(r) v(r)}{r}\,\\
 \varepsilon_{QM}(r)/3&=&\left(\psi\right)^\dagger(\bfvec{r})\, \beta \,M_q\,\left(\psi\right)(\bfvec{r}) =M_q\,\left(u^2(r) -v^2(r)\right)\\
 \varepsilon_{C}(r)/3&=&\left(\psi\right)^\dagger(\bfvec{r})\,W_C(\bfvec{r})\,\left(\psi\right)(\bfvec{r}) = W_C(r)\,\left(u^2(r) +v^2(r)\right),\\
 -p_{C}(r)/3&=&-\left(\psi\right)^\dagger(\bfvec{r})\, \bfvec{}\cdot\bfvec{\nabla}W_C(\bfvec{r})\,\left(\psi\right)(\bfvec{r}) =-r\, W'_C(r)\,\left(u^2(r) +v^2(r)\right).
\end{eqnarray}
For the physical nucleon state, $|N: \bfvec{0}_P\rangle\,\equiv\,|N(\bfvec{P}_N=0)\rangle$, the various contributions to the quark energy densities and pressure distributions, are  given by:
 \begin{eqnarray}
\varepsilon_{QK}(r)&=&3\int\frac{V d^3\Delta}{(2\pi)^3}\, \langle \bfvec{\Delta}_N/2|\,q^\dagger(\bfvec{r})\,(-i\bfvec{\alpha}\cdot\vec\nabla_{\bfvec{r}})\,q(\bfvec{r})|-\bfvec{\Delta}_N/2\rangle\nonumber\\
&=&\int\frac{d^3\Delta}{(2\pi)^3}\, e^{-i\,\bfvec{\Delta} \cdot\bfvec{r}}\, e^{5 b^2\Delta^2/24}\,\int d^3x\,e^{i\,\bfvec{\Delta} \cdot\bfvec{x}} \,\left(\psi\right)^\dagger_E({\bf{x}})\,(-i\bfvec{\alpha}\cdot\vec\nabla_{\bfvec{x}})\left(\psi\right)_E({\bf{x}})\nonumber\\
&=&\int\frac{d^3\Delta\,d^3x}{(2\pi)^3}\, e^{5 b^2\Delta^2/24}\,j_0(\Delta r)\,j_0(\Delta x)\,
\left(\psi^\dagger_E \,(-i\bfvec{\alpha}\cdot\vec\nabla_{\bfvec{x}})\,\psi_E\right)(x)\nonumber\\
&\equiv& E_{QK}\,\rho_K(r) \label{EQKW},\\
\varepsilon_{QM}(r)&=&3\int\frac{V d^3\Delta}{(2\pi)^3}\, \langle \bfvec{\Delta}_N/2|\,q^\dagger(\bfvec{r})\,(\beta\,M_q)\,q(\bfvec{r})|-\bfvec{\Delta}_N/2\rangle\nonumber\\
&=&\int\frac{d^3\Delta}{(2\pi)^3}\, e^{-i\,\bfvec{\Delta} \cdot\bfvec{r}}\, e^{5 b^2\Delta^2/24}\,\int d^3x\,e^{i\,\bfvec{\Delta} \cdot\bfvec{x}} \,\left(\psi\right)^\dagger_E({\bf{x}})\,(\beta\,M_q)\left(\psi\right)_E({\bf{x}})
\nonumber\\
&=&\int\frac{d^3\Delta\,d^3x}{(2\pi)^3}\, e^{5 b^2\Delta^2/24}\,j_0(\Delta r)\,j_0(\Delta x)\,
\left(\psi^\dagger_E \,(\beta\,M_q)\,\psi_E\right)(x)\nonumber\\
&\equiv& E_{QM}\,\rho_S(r), 
\label{EQMW}
\end{eqnarray}
\begin{eqnarray}
\varepsilon_C(r)&=&   \int\frac{V d^3\Delta}{(2\pi)^3}\, \langle \bfvec{\Delta}_N/2|\,\mathcal{H}_C(\bfvec{r}) \,|-\bfvec{\Delta}_N/2\rangle= W_C(r)\,\int\frac{V d^3\Delta}{(2\pi)^3}\, \langle \bfvec{\Delta}_N/2|\,q^\dagger(\bfvec{r})\,q(\bfvec{r})|-\bfvec{\Delta}_N/2\rangle\nonumber\\
 &=& 3\,W_C(r)\,\rho_B(r),\label{EQCW}
\end{eqnarray}
\begin{eqnarray}
- p_C(r)&=& -3\,  \int\frac{V d^3\Delta}{(2\pi)^3}\, \langle \bfvec{\Delta}_N/2|\,\mathcal{P}_C(\bfvec{r})\,| -\bfvec{\Delta}_N/2\rangle=- r W'_C(r)\,\int\frac{V d^3\Delta}{(2\pi)^3}\, \langle \bfvec{\Delta}_N/2|\,q^\dagger(\bfvec{r})\,q(\bfvec{r})|-\bfvec{\Delta}_N/2\rangle\nonumber\\
 &=&- \, r\,W'_C(r)\,\rho_B(r).    
\end{eqnarray}
\begin{eqnarray}
\rho_B(r)&=&\int\frac{V d^3\Delta}{(2\pi)^3}\, \langle \bfvec{\Delta}/2|\,q^\dagger(\bfvec{r})\,q(\bfvec{r})|N(-\bfvec{\Delta}/2)\rangle\nonumber\\
&=&\int\frac{d^3\Delta}{(2\pi)^3}\, e^{-i\,\bfvec{\Delta} \cdot\bfvec{r}}\, e^{5 b^2\Delta^2/24}\,\int d^3x\,e^{i\,\bfvec{\Delta} \cdot\bfvec{x}} \,\left(\psi\right)^\dagger_E({\bf{x}})\,\left(\psi\right)_E({\bf{x}})\\
&=&\int\frac{d^3\Delta\,d^3x}{(2\pi)^3}\, e^{5 b^2\Delta^2/24}\,j_0(\Delta r)\,j_0(\Delta x)\,
\left(\psi^\dagger_E \psi_E\right)(x),\label{EQRHOB}
\end{eqnarray}
with the normalization condition:
\begin{equation}
\int d^3 r \, \varepsilon_{QK}(r) =E_{QK},\qquad\qquad  \int d^3 r \, \varepsilon_{QM}(r) =E_{QM},\qquad\qquad\int d^3 r  \rho_B(r) =1.
\end{equation}
%%%%%%%%%%%%%%%%%%%%%%%%%%%%%%%%%%%%%%%%%%%%%%%%%%%%%%%%%%%%%%%%%%%%%%%%%%%%%%%%%%%%%%%%%%%%%%%%%%%%
\subsection{Pionic contribution to the energy density and pressure}
For the localized CCM static bag, $|N: \bfvec{0}_X\rangle\,\equiv\,|N(\bfvec{X}_N=0)\rangle$, the various contributions to the  energy densities and pressure distributions of pionic origin, are  given by:
\begin{eqnarray}
 \varepsilon_{\pi M}(r)&=&\frac{3}{2} \left(\frac{1}{2 F_\pi(\mathcal{S})}\right)^2 \int\frac{d^3 k}{(2\pi)^3} \frac{d^3 q}{(2\pi)^3}\,j_1(kr)\,j_1(qr)\,\Gamma(k) \,\Gamma(q)\, \frac{M^2_\pi(\mathcal{S})\,k\,q}{2\omega_k\omega_q}\nonumber\\
&& \bigg[\frac{1}{(\omega_k+\omega_q)(\omega_k+\bar{\epsilon}_{Nkq})}+\frac{1}{(\omega_k+\omega_q)(\omega_q+\bar{\epsilon}_{Nkq})}+\frac{1}{(\omega_k+\bar{\epsilon}_{Nkq})(\omega_q+\bar{\epsilon}_{Nkq})}
\nonumber\\
&&+\frac{32}{25}\,\bigg(\frac{1}{(\omega_k+\omega_q)(\omega_k+\bar{\epsilon}_{\Delta kq})}+\frac{1}{(\omega_k+\omega_q)(\omega_q+\bar{\epsilon}_{\Delta kq})}+\frac{1}{(\omega_k+\bar{\epsilon}_{\Delta kq})(\omega_q+\bar{\epsilon}_{\Delta kq})}\bigg)\bigg]\label{DPIM}
\end{eqnarray}
\begin{eqnarray}
 \varepsilon_{\pi K}(r)&=&\frac{3}{2} \left(\frac{1}{2 F_\pi(\mathcal{S})}\right)^2 \int\frac{d^3 k}{(2\pi)^3} \frac{d^3 q}{(2\pi)^3}\,\left(\frac{1}{3}j_0(kr)\,j_0(qr)\,+\,\frac{2}{3}j_2(kr)\,j_2(qr)\right)\,\Gamma(k) \,\Gamma(q)\, \frac{k^2\,q^2}{2\omega_k\omega_q}\nonumber\\
&& \bigg[\frac{1}{(\omega_k+\omega_q)(\omega_k+\bar{\epsilon}_{Nkq})}+\frac{1}{(\omega_k+\omega_q)(\omega_q+\bar{\epsilon}_{Nkq})}+\frac{1}{(\omega_k+\bar{\epsilon}_{Nkq})(\omega_q+\bar{\epsilon}_{Nkq})}
\nonumber\\
&&+\frac{32}{25}\,\bigg(\frac{1}{(\omega_k+\omega_q)(\omega_k+\bar{\epsilon}_{\Delta kq})}+\frac{1}{(\omega_k+\omega_q)(\omega_q+\bar{\epsilon}_{\Delta kq})}+\frac{1}{(\omega_k+\bar{\epsilon}_{\Delta kq})(\omega_q+\bar{\epsilon}_{\Delta kq})}\bigg)\bigg]\label{DPIK}
\end{eqnarray}
\begin{eqnarray}
 \varepsilon_{\pi c}(r)&=&-\frac{3}{2} \left(\frac{1}{2 F_\pi(\mathcal{S})}\right)^2 \int\frac{d^3 k}{(2\pi)^3} \frac{d^3 q}{(2\pi)^3}\,j_1(kr)\,j_1(qr)\,\Gamma(k) \,\Gamma(q)\, \frac{k\,q\,(\omega^2_k+\omega^2_q)}{4\omega_k\omega_q}\nonumber\\
&& \bigg[\frac{1}{(\omega_k+\omega_q)(\omega_k+\bar{\epsilon}_{Nkq})}+\frac{1}{(\omega_k+\omega_q)(\omega_q+\bar{\epsilon}_{Nkq})}+\frac{1}{(\omega_k+\bar{\epsilon}_{Nkq})(\omega_q+\bar{\epsilon}_{Nkq})}
\nonumber\\
&&+\frac{32}{25}\,\bigg(\frac{1}{(\omega_k+\omega_q)(\omega_k+\bar{\epsilon}_{\Delta kq})}+\frac{1}{(\omega_k+\omega_q)(\omega_q+\bar{\epsilon}_{\Delta kq})}+\frac{1}{(\omega_k+\bar{\epsilon}_{\Delta kq})(\omega_q+\bar{\epsilon}_{\Delta kq})}\bigg)\bigg]\nonumber\\
&&+\frac{3}{2} \left(\frac{1}{2 F_\pi}\right)^2 \int\frac{d^3 k}{(2\pi)^3} \frac{d^3 q}{(2\pi)^3}\,j_1(kr)\,j_1(qr)\,\Gamma(k) \,\Gamma(q)\,\frac{k\,q}{2} \nonumber\\&&\bigg[\frac{1}{\omega_k}\frac{1}{\omega_k+\bar{\epsilon}_{Nkq}} + \frac{1}{\omega_q}\frac{1}{\omega_q+\bar{\epsilon}_{Nkq}} _,+\,\frac{32}{25}\,\left(\frac{1}{\omega_k}\frac{1}{\omega_k+\bar{\epsilon}_{\Delta kq}}+\frac{1}{\omega_q}\frac{1}{\omega_q+\bar{\epsilon}_{\Delta kq}}\right) \bigg].\label{DPIC}
\end{eqnarray}
\begin{eqnarray}
 \varepsilon_{\pi Q}(r)&=&-3\left(\frac{1}{2 F_\pi(\mathcal{S})}\right)^2\int\frac{d^3 k}{(2\pi)^3}   \,k^2\,\,\left(j_0(kr)\,\tilde{\Gamma}_0(r)\,+\,j_2(kr)\,\tilde{\Gamma}_2(r)\right)\,\Gamma(k)\nonumber\\
 &&\bigg[\frac{1}{\omega_k}\frac{1}{\omega_k+\epsilon_{N {\bf k}}} +\frac{32}{25}\,\frac{1}{\omega_k}\frac{1}{\omega_k+\epsilon_{\Delta{\bf
k}}} \bigg].\label{DPIQ}   
\end{eqnarray}
with:
\begin{eqnarray}
 \tilde{\Gamma}_0(r)&=&\int \frac{d^3 t}{(2\pi)^3}\, j_0(tr)\,\Gamma_0(t)\\
 \tilde{\Gamma}_2(r)&=&\int \frac{d^3 t}{(2\pi)^3}\, j_2(tr)\,\Gamma_2(t)\\
 \Gamma_0(t)&=& P(t)\,\frac{5}{3}\int d^3x\,\left(u^2(x)-\frac{v^2(x)}{3}\right)\\
 \Gamma_2(t)&=& P(t)\,\frac{5}{3}\left(-\frac{4}{3}\int d^3x\,\,j_2(tx) \,v^2(x)\right)\\
 \Gamma(k)&=& P(k)\,\frac{5}{3}\int d^3x\,\left[j_0(kx)\left(u^2(x)-\frac{v^2(x)}{3}\right)-\frac{4}{3}\,j_2(kx) \,v^2(x)\right].
\end{eqnarray}
For the physical nucleon state, $|N: \bfvec{0}_P\rangle\,\equiv\,|N(\bfvec{P}_N=0)\rangle$, one has formally  the same expressions for the various contribution to the  energy densities of pionic origin but with the $b$ parameter replaced by $b_E=\sqrt{3/2}\,b$, and the form factor $\Gamma(k)$ replaced by $\Gamma(k)\exp(5 b^2 k^2/24)$, namely
\begin{equation}
 \Gamma(k)=  e^{5 b^2k^2/24}\,P(k)\,\frac{5}{3}\int d^3x\,\left[j_0(kx)\left(u_E^2(x)-\frac{v_E^2(x)}{3}\right)-\frac{4}{3}\,j_2(kx) \,v_E^2(x)\right].   
\end{equation}

%%%%%%%%%%%%%%%%%%%%%%%%%%%%%%%%%%%%%%%%%%%%%%%%%%%%%%%%%%%%%%%%%%%%%%%%%%%%%%%%%%%%%%%%%%%%%%%%%%%%%%%ù%%%%%%%%%%%%%%%%%%%%%%%%%%%%%%%%%%%%%%%%%%%%%%%%%%%%%%%%%%%%%%%%%%%%ù


\begin{thebibliography}{999}
\bibitem{Nucleon-stability-I} G. Chanfray, H.Hansen,  B. K. Pradhan, companion paper: "Mechanical properties of the nucleon in the chiral confining model. I - formal developments"; arXiv : 2606.03588 [nucl-th].
\bibitem{SerotWalecka1986} B. D. Serot, J. D. Walecka, The Relativistic Nuclear Many-Body Problem, Advances in Nucl. Phys. 16, 1 (1986).
\bibitem{Walecka1997} B. D. Serot, J. D. Walecka, Int. J. Mod. Phys. E 16, 15 (1997).
\bibitem{Chanfray2001} G. Chanfray, M. Ericson and P. A. M. Guichon, Phys. Rev. C 63, 055202 (2001).
\bibitem{Martini2006} G. Chanfray, D. Davesne, M. Ericson and M. Martini, Eur. Phys. J. A 27, 191 (2006).
\bibitem{Boguta83} J. Boguta, Phys. Lett. 120B, 34 (1983).
\bibitem{KM74} A. K. Kerman and L. D. Miller in ``Second High Energy Heavy Ion Summer Study'',LBL-3675, 1974. 
\bibitem{BT01} W. Bentz, A. W. Thomas, Nucl. Phys. A 696 (2001) 138.
\bibitem{C03} G. Chanfray,  Nucl. Phys. A 721, 76c (2003).
\bibitem{Chanfray2005} G. Chanfray, M. Ericson, Eur. Phys. J. A 25, 151 (2005).
\bibitem{Chanfray2007} G. Chanfray, M. Ericson, Phys. Rev. C 75, 015206 (2007).
\bibitem{Ericson2007} M. Ericson, G. Chanfray, Eur. Phys. J. A 34, 215 (2007).
\bibitem{Massot2008} E. Massot, G. Chanfray, Phys. Rev. C 78, 015204 (2008).
\bibitem{Massot2009} E. Massot, G. Chanfray, Phys. Rev. C 80, 015202 (2009).
\bibitem{Guichon1988} P. A. M. Guichon, Phys. Lett. B 200, 235 (1988).
\bibitem{Guichon96} P. A. M Guichon, K. Saito, E. Rodionov and A. W. Thomas, Nucl. Phys. A 601, 349 (1996).
\bibitem{Guichon2004} P. A. M. Guichon, A. W. Thomas, Phys. Rev. Lett. 93, 132502 (2004). 
\bibitem{Guichon2006}P. A. M. Guichon, H. H. Matevosyan, N. Sandulescu and A. W. Thomas, Nucl. Phys. A 772, 1 (2006).
\bibitem{Stone}R. Stone, P. A. M. Guichon, P. G. Reinhard and A. W. Thomas,  Phys. Rev. Lett. 116, 092511 (2016).
\bibitem{QMCreview} P. A. M Guichon, J. R. Stone and A. W. Thomas, Progr. Part. Nucl. Phys. 100, 262 (2018).
\bibitem{Massot2012} E. Massot, J. Margueron and G. Chanfray, EPL 97, 39002 (2012).
\bibitem{Rahul} R. Somasundaram, J. Margueron, G. Chanfray and H. Hansen, Eur. Phys. J. A 58, 5 (2022).
\bibitem{Cham1} M. Chamseddine, J. Margueron, G. Chanfray, H. Hansen and  R. Somasundaram,   Eur. Phys. J. A 59 (2023) 8, 177; arXiv:2304.01817 [nucl-th].
\bibitem{Chanfray2023} G. Chanfray, H. Hansen and J. Margueron, Eur. Phys. J. A 59, 11 (2023); arXiv: 2304.01036 [nucl-th].
\bibitem{Universe} G. Chanfray, M. Ericson and M. Martini, Universe 9 (2023) 7, 316; arXiv: 2307.03484 [nucl-th].
\bibitem{Chanfray2024} G. Chanfray, Eur. Phys. J. A 60, 1 (2024); arXiv: 2310.19532 [nucl-th].
\bibitem{Cham2} M. Chamseddine, J. Margueron, H. Hansen and  G. Chanfray,   
Eur. Phys. J. A 60 (2024) 6, 137; arXiv: 2401.10100 [nucl-th].
\bibitem{LTY03}D. B. Leinweber, A. W. Thomas and R. D. Young, arXiv: 0302020 [hep-lat].
\bibitem{LTY04}D. B. Leinweber, A. W. Thomas and R. D. Young, Phys.
Rev. Lett. 92, 242002 (2004).
\bibitem{TGLY04} A. W. Thomas, P. A. M. Guichon, D. B.  Leinweber and R. D. Young
Progr. Theor. Phys. Suppl. 156 (2004) 124; arXiv: 0411014 [nucl-th].
\bibitem{AALTY10} W. Armour, C.R. Alton,D. B. Leinweber, A. W. Thomas and R. D. Young, Nucl. Phys. A 840, 97 (2010).
\bibitem{TB1} J. I. Fujita and H. Miyazawa, Prog. Theor. Phys. 17, 360 (1957).
\bibitem{TB2} G.  Brown and A. M Green, Nucl. Phys. A 137, 1 (1969). 
\bibitem{TB3} S. A. Coon, M. D. Scadron, P. C. McNamee, B. R. Barrett, D. W. E. Blatt. and B. H. J. McKeller, Nucl. Phys. A 317, 242 (1979).
\bibitem{TB4} H. T. Coelho, T. K. Das, and M. R. Robilotta, Phys. Rev. C 28, 1812 (1983). 
\bibitem{TB44} M. R. Robilotta and H. T. Coelho, Nucl.
Phys. A 460, 645 (1986).
\bibitem{CEFT} R. Machleidt and D. R. Entem,   Phys. Rep. 503, 1 (2011); arXiv: 1105.2919 [nucl-th].
\bibitem{Chanfray2011} G. Chanfray, M. Ericson, Phys. Rev. C 83, 015204 (2011).
\bibitem{Chan} Lai-Him Chan, Phys. Rev. Lett. 57, 1199 (1986).
\bibitem{Universe2025} G. Chanfray, M. Ericson, H. Hansen, J. Margueron, and M. Martini, Symmetry 17, 313 (2025), arXiv:2501.10177 [nucl-th].
\bibitem{Simonov1997} Yu. A. Simonov, Phys. At. Nucl. 60, 2069 (1997).
\bibitem{Simonov1998}Yu. A. Simonov, Few-BodySyst. 25, 45 (1998).
\bibitem{Tjon2000} Yu. A. Simonov, J. A. Tjon, Phys. Rev. D 62, 014501 (2000).
\bibitem{Simonov2002} Yu. A. Simonov, J. A. Tjon, J. Weda, Phys. Rev. D 65, 094013 (2002). 
\bibitem{Simonov2002a}Yu. A. Simonov, Phys. Rev. D 65, 094018 (2002).
\bibitem{Simonov-light} Yu. A. Simonov, Phys. At.Nucl. 67, 846 (2004), arXiv: 0302090 [hep-ph].
\bibitem{light1}Yu. A. Simonov, Phys. At. Nucl. 67, 1027 (2004), arXiv: 0305281 [hep-ph].
\bibitem{light2} Yu. A. Simonov, Int. Mod. Phys. A 31, 165016 (2016), arXiv:1509.06930 [hep-ph].
 \bibitem{Digiacomo} A. Di Giacomo, H. G. Dosch, V. I. Shevchenko and Yu .A. Simonov, Phys. Rep. 372,  319 (2002).
 \bibitem{Simonov2019} Yu. A. Simonov, Phys. Rev. D 99, 056012 (2019).
\bibitem{Gastao} G. Krein, Eur. Phys. J. A. 18, 511 (2003); M. E. Bracco, G. Krein, M. Nielsen, Phys. Lett. B 432, 258 (1998).
\bibitem{Xiong} W. Xiong et al., Nature (London) 575, 147 (2019).
 \bibitem{Brown} G. E. Brown, Nucl. Phys. A 358, 39c (1981).
\bibitem{Pirner} J. De Kam, H. J. Pirner, Nucl. Phys. A 389, 640 (1982).
\bibitem{KLE} S. P. Klevansky, Review of Modern Physics 64, 649 (1992).
\bibitem{Lorce2021} C. Lorcé, A. Metz, B. Pasquini and S. Rodini, jheph11, 121 (2021).
\bibitem{vonLaue}M. von Laue, Ann. Phys. (Leipzig) 340, 524 (1911).
\bibitem{Polyakovor} M. V. Polyakov, Phys. Lett. B555, 57 (2003).
\bibitem{Goeke} K. Goeke, J. Grabis, J. Ossmann, M.V. Polyakov, P. Schweitzer, A. Silva, and D. Urbano, Phys. Rev. D75, 094021 (2007).
 \bibitem{Polyakov} M.V. Polyakov, P. Schweitzer, Int.J.Mod.Phys. A33,  1830025 (2018), arXiv:1805.06596 [hep-ph].
 \bibitem{Lorce2019} C. Lorcé, H. Moutarde, A. P. Trawinski,  Eur. Phys. J. C,  79 (2019) arXiv:1810.09837 [hep-ph].
 \bibitem{RGS} R. G. Sachs, Phys. Rev. 126, 2256 (1962).
 arXiv: 2511.1167 [nucl-th].
 \bibitem{DVCS} V.D. Burkert, L. Elouadrhiri, F.X. Girod, Nature 557  7705, 396 (2018).
\bibitem{MESPROD} L. Favart, M. Guidal, T. Horn, P. Kroll, Eur. Phys. J. A 52, 158 (2016); arXiv:1511.04535.
\bibitem{Wigner}M. Hillery, R.F. O’Connell, M.O. Scully, E.P.Wigner, Phys. Rep. 106, 121 (1984).
\bibitem{Fuku}K. Fukushima, T. Kojo and W. Weise, Phys. Rev. D 102, 096017 (2020); arXiv: 2008.08436 [hep-ph]. 
 \bibitem{Pradhan} B. K. Pradhan, G. Chanfray, H. Hansen, J. Margueron, Eur.Phys.J.A 62 (2026) 8, 178, arXiv: 2511.11167[nucl-th].
 %\bibitem{Diakonov} D. Diakonov et al, Nucl. Phys. B480 (1996). 


\end{thebibliography}
\end{document}